# Dynamics of Rotation of Super-Earths

## N. CALLEGARI JR. and A. RODRÍGUEZ

**Abstract** We numerically investigate the dynamics of rotation of several close-in terrestrial exoplanet candidates. In our model, the rotation of the planet is disturbed by the torque of the central star due to the asymmetric equilibrium figure of the planet. We model the shape of the planet by a Jeans spheroid. We use surfaces of section and spectral analysis to explore numerically the rotation phase space of the systems adopting different sets of parameters and initial conditions close to the main spin-orbit resonant states. One of the parameters, the orbital eccentricity, is critically discussed here within the domain of validity of orbital circularization timescales given by tidal models. We show that, depending on some parameters of the system like the radius and mass of the planet, eccentricity etc, the rotation can be strongly perturbed and a chaotic layer around the synchronous state may occupy a significant region of the phase space. 55 Cnc e is an example.

# 1 Introduction

## 1.1 Motivation

Many studies on the physics of exoplanets assume an equilibrium state for the rotation of the planet. Following we provide a brief list of studies of different

N. CALLEGARI JR.
Instituto de Geociências e Ciências Exatas, Unesp - Univ Estadual Paulista, Av. 24-A, 1515, CEP 13506-900, Rio Claro/SP/Brazil.
Tel.: 55-19-3526-9132
E-mail: calleg@rc.unesp.br
A. RODRÍGUEZ
Departamento de Astronomia, Instituto de Astronomia, Geofísica e Ciências Atmosféricas - Universidade de São Paulo, Rua do Matão, 1226, CEP 05508-090, São Paulo/SP/Brazil.
Tel.: 55-11-3091-2704
E-mail: adrian@astro.iag.usp.br

themes related to exoplanets which assume in their models that the planet is tidally locked: measurement of the rotation and oblateness of transiting giant exoplanets (e.g. Seager and Hui 2002; Barnes et al. 2003; Carter and Winn 2010); generation of magnetic fields in exoplanets (e.g. Zuluaga and Cuartas-Restrepo 2012; see also Lammer et al. 2009 and references therein); atmosphere and climate of exoplanets (e.g. Lammer et al. 2008, Dobrovolskis 2007, Dobrovolskis 2009, Showman and Polvani 2011, Castan and Menon 2011); secular orbital evolution due to tidal effects (e.g. Jackson et al. 2008a; Rodríguez et al. 2011); tidal heating (e.g. Levrard et al. 2007, Wisdom 2008, Jackson et al. 2008b,c); interior structure of exoplanets (e.g. Ragozzine and Wolf 2009, Batygin et al. 2009; Kramm et al. 2011).

It is well known that the detected exoplanets generally have unusual orbits when compared to planets in the Solar System. For instance, many planets are much closer to their stars than Mercury is to the Sun. Some of these planets, the so-called close-in planets, also have eccentric orbits (see Table I). Thus, we can ask if the strength of the torque of the star in their rotational motion may be important. The main task here is to investigate this question. Many studies have been published on the global dynamics of rotation of fictitious exoplanets which explore, analytically and/or numerically, the rotation phase space (e.g. Kitiashvili and Alexander 2008; Celletti and Voyatzis 2010). Our contribution here is to provide details on the dynamics of rotation and stability of several known Earth-like candidates, in the vicinity of the synchronism and other spin-orbit resonant states.

In this study, we focus on the conservative model, where we do not investigate the effects on the spin rate of the planet due to dissipative forces, like those associated with tides. Therefore we consider here only the gravitational torque due to the non-sphericity of the planet. However, our investigation can be useful in the interpretation of results in which tidal forces are taken into account in the rotational history of the planet rotation (e.g. Rodríguez et al. 2012).

## 1.2 The model

Several aspects of the physics of rotation of close-in exoplanets can be understood using the 1.5-degree-of-freedom dynamical system often applied in studies of physical libration in rotation of regular satellites of the Solar system, the Moon and Mercury (e.g. Goldreich and Peale 1966):

$$\ddot{\theta} = \frac{3}{2}\left(\frac{B-A}{C}\right)\frac{Gm_0}{r^3}\sin 2(f-\theta), \tag{1}$$

$$\dot{f} = \frac{\sqrt{G(m_0+m)a(1-e^2)}}{r^2}, \qquad r = \frac{a(1-e^2)}{1+e\cos f}, \tag{2}$$

where: $A < B < C$ are the moments of inertia around the principal axis of the planet; $G$ is the gravitational constant; $m_0$, $m$ are the masses of the central star and the planet, respectively; $r$, $a$, $e$, $f$ are the planet-star distance, semi-major axis, eccentricity and true anomaly, respectively. $\theta$ is an angle measured from an inertial line such that $\dot{\theta}$ is the angular velocity of rotation around the axis with greatest moment of inertia, which is assumed to be perpendicular to the orbital plane. The mutual perturbations between the bodies belonging to multiple systems are neglected in this paper, and the motion of the system is governed by the laws of the two-body problem[1]. The equator of the planet is non-circular (elliptical), and therefore, in the model, the rate of rotation of the secondary body is disturbed by the torque of the star on the non-spherical shape of the planet.

A basic assumption in the application of the model (1) is that the planet has a permanent quadrupole structure. Examples of bodies with this physical property in the solar system are the terrestrial planets and the regular satellites, all of which are, in spite of the details of their interior structures, solid-like bodies. In the case of exoplanets, CoRoT-7b, Kepler-10b and 55 Cnc-e are examples of Super-Earths that are probably solid-like (Léger et al. 2009, Batalha et al. 2011, Valencia 2011, Gillon et al. 2012). They are transiting planets which physical parameters which are the subject of much research. For instance, the mass of CoRoT-7b has been estimated by several teams which have obtained different results (see Ferraz-Mello et al. 2011).

It is still difficult to be sure about the structure of probably Earth-like exoplanets at the current status (see Valencia 2011, Léger et al. 2011, for the case of CoRoT-7b). In the case of exoplanets, important new developments have contributed to existing knowledge on planetary structure (see Baraffe et al. 2010, Valencia 2011, Léger et al. 2011, and references therein). In particular, Valencia et al. (2007a) show that, for a determined value of the mass of a rock planet in the interval $m < 10M_E$ ($M_E$ is the Earth mass), its radius must be smaller than a determined value which depends on $m$. The limits of the radius for a given mass are the following: 6600; 8600; 10,440; 11,600 and 12,200 km for 1, 2.5, 5, 7.5 and 10 $M_E$, respectively. So, if a planet with $m < 10M_E$ has a radius greater than its corresponding value in the above interval, its composition can include significant quantities of material with density less than that of the rock component, like water for instance. In that case, a permanent deformation is not, in principle, guaranteed. In the same sense, Hot-Jupiters, like the solar system's jovian planets, are probably composed mainly of gas, and we did not include them in the present study because additional discussions is needed in order to consider a fixed structure for these bodies (see Levrard et al. 2007, Rodríguez et al. 2012).

Our main goal in this work, i.e., the application of the model (1) for solid-like planets, has one main limitation: total ignorance regarding the quantity $\frac{B-A}{C}$ which appears in the amplitude of the perturbation of the planet rotation. In Section 2, we propose a method to estimate the value of $\frac{B-A}{C}$ of Super-Earths with known values of radius and mass. In Section 2, we also show how to infer $\frac{B-A}{C}$ as a function of several parameters (e.g., mass, distance, radius etc) of systems containing planets for which the transiting measurement, and therefore the radius, are not known.

---

[1] In this work only the prograde direction of both, orbital and rotational motion, are considered. However, exoplanets in retrograde orbits with respect to the rotation axis of the star have been detected (e.g. Hébrard et al. 2011, and references therein). Retrograde rotation is also a possible equilibrium state (Correia et al. 2008). The study of spin-orbit resonances in retrograde motion would be interesting.

### 1.3 Methodology

Wisdom et al. (1984) study solutions of the Equations (1,2) through the computation of surfaces of sections. They numerically show the possibility of chaotic rotation around the synchronous and other spin-orbit states for non-spherical satellites orbiting their planets in compact and eccentric orbits. We use Wisdom's methodology and apply it to the rotation dynamics of exoplanets. In Section 3, we show the main results of numeric integrations of Equations (1,2) for several planets listed in Table I. In some cases, we also compare the results obtained with surfaces of sections with dynamical maps based in spectral analysis of the solutions (e.g. Michtchenko and Ferraz-Mello 2001).

Our simulations cover wide ranges of the free parameters including moments of inertia ratio and orbital eccentricity. We also give the range of some other parameters (like planetary radius, mass) which would correspond to solid-like planets in view of structure models of exoplanets (e.g. Valencia 2007a,b; 2011). Thus, we discuss the dynamics of planetary rotation with our model in large sets of initial conditions and appropriate parameters corresponding to solid-like planets.

We consider planets with orbital period $P < 33$ days and mass $m < 15 M_E$. Table I shows the data of several Super-Earths considered here. Inspection of Table I helps us to see some common properties of these planets. For instance, there are several single-planets with published value of $e > 0$, even with very small values of orbital period $P < 7$ days. As we see in Appendix A (see below), the existence of non-circular orbits for single short-period planets can be in conflict with the classical results of orbital circularization of close-in orbits due to tidal effects[2].

### 1.4 Orbital eccentricity

As we will see in Section 3, the planets with more rich rotational dynamics are those which move in eccentric orbits. In fact, orbital eccentricity plays an important role, since in the case of non-circular orbit, the amplitude and argument of the sine in (1) undergo temporal variation. However, the existence of eccentric close-in orbits in single-planet systems cannot be guaranteed. In fact, an initial eccentric orbit would be tidally damped, evolving to circularization in timescales depending on physical parameters of the interacting pair and also on the initial values of the elements (see Dobbs-Dixon et al. 2004, Ferraz-Mello et al. 2008, Rodríguez and Ferraz-Mello 2010, Rodríguez 2010). In Appendix A we calculate, using averaged equations, the circularization timescale of the Super-Earth CoRoT-7b, adopting different dissipative parameters. Our results, which are in agreement with previous ones, show that tidal evolution of systems with close-in companions with orbital period less than one week may be responsible for the circularization of the orbits.

Thus, the values of the eccentricities of some close-in *single-planets* listed in Table I are probably not determined very well by the current methods of orbital fitting, and/or some additional mechanism may explain the non-null values. For instance, under the hypothesis of an additional undetected resonant companion, a large value of eccentricity of a single planet can be in fact close to zero, a result that shows a degeneracy of the problem of orbital determination of the system (Anglada-Escudé et al. 2010). Tadeu dos Santos et al. (2012), in a paper written mainly to describe the dynamics and the determination of the orbital elements of Gl 581g, show degeneracy in the determination of the eccentricity of all members of the Gl 581 system. In this case, degeneracy means that, depending on the planetary architecture adopted in the model (e.g. number of planets), different solutions are possible with the same statistical significance for different values of eccentricity.

---

[2] Note however that the timescale for orbital circularization decreases with the stellar mass (e.g. Rodríguez and Ferraz-Mello 2010). Thus, some single-planets orbiting small stars may have large circularization timescales (for example, GJ 1214b, among others).

The existence of exterior companions can work as a dynamical mechanism which may explain the non-null eccentricity values of close-in planets of *multi-planetary* systems. Spiegel et al. (2010) suggest that terrestrial planets at 1 AU can be highly perturbed by an outer giant planet, which may excite large variations of eccentricity of the inner planet. At such distances, however, tidal effects may be negligible (Appendix A), and eccentricity may be large. However, previous studies have shown that the combination of secular interaction and tides raised on the inner planet by the star results in the circularization of both orbits (see Mardling 2007; Rodríguez et al. 2011). Indeed, depending on the mass ratio of the planet, the perturbation of the companion acts to increase the circularization timescale of the inner orbit. Hence, the explanation for the non-null eccentricities of close-in planets remains open to discussion (see Matsumura et al. 2008, Correia et al. 2010, Correia et al. 2012).

Therefore, notwithstanding the above discussion, in the present study, orbital eccentricity is considered an open parameter in our calculations. The reported values, listed in Table I with the error bars, will also be taken into account for reference.

## 2 The Prolateness

The first quantity which we must have in hand in order to study the system (1-2) is the ratio $\frac{B-A}{C}$. Here we propose a method to estimate the value of $\frac{B-A}{C}$ as a function of several parameters of the system, including the mass of the star, the mass of the planet, semi-major axis and orbital eccentricity, and the radius of the planet.

Consider a homogeneous *non-viscous*[3] rotating planet orbiting a star with angular velocity $\Omega$. The equilibrium figure of a rotating body (with axes a, b and c) under the action of the tide generating force of the central star, lying in the equatorial body's plane, is given by a Roche ellipsoid in which a > b > c (see Chandrasekhar 1969). The smaller axis is perpendicular to the largest axis, which is pointed in the direction of the central body in the case of synchronous rotation. In this case, we have the following expression for the deformation of the equator, or the *prolateness*, hereafter denoted by $\epsilon$:

---

[3] In this hypothesis we are neglecting the dynamic tidal torque raised by the central star which also affects the planet rotation. However, several planets which we study here are so close to their star that tidal effects may be important, mainly in the case of eccentric orbits. Appendix B presents one case in which the tidal torque is taken into account and applied to CoRoT-7b super-Earth planet.

$$\epsilon = \frac{\mathrm{a} - \mathrm{b}}{\mathrm{a}} = \frac{15}{4} \frac{m_0}{m} \left(\frac{R}{r}\right)^3 \left(1 - \frac{5(m_0 + m)}{2m} \left(\frac{\Omega}{n}\right)^2 \left(\frac{R}{r}\right)^3\right)^{-1} \tag{3}$$

(e.g. Ferraz-Mello et al. 2008). Table II shows the values of $\epsilon$ considering $\Omega = n$ for several transiting Super-Earths.

*Table I:* **Data of planets with orbital period $P < 33$ days (except Kepler-11f) and mass $m < 15 M_E$, where $M_E$ is the Earth mass. The list is sorted in increasing value of the orbital period. The data have been taken from http://exoplanet.eu/index.php except in other cases indicated. The other units are the solar mass ($M_{SUN}$) and the Astronomical Unit (AU). [a]: Ferraz-Mello et al. (2011); [b]: Planet present in multiple system; [c]: See also Section 3.2.1.**

| Planet | $m$: Mass ($M_E$) | $m_0$: Star's mass ($M_{SUN}$) | $P$: Orbital Period (day) | $a$: semi-major axis (AU) | $e$: eccentricity |
|--------|------|------|------|------|------|
| KOI-55b | 0.445 | 0.496 | 0.2401 | 0.006 | ?[b] |
| KOI-961c | 1.907 | 0.13 | 0.4533 | 0.006 | ?[b] |
| KOI-55c | 0.667 | 0.496 | 0.3429 | 0.0076 | ?[b] |
| 55 Cnc e | 8.58 | 0.905 | 0.7365 | 0.0156 | $< 0.06^{b}$ |
| Kepler-10b | 4.55 | 0.895 | 0.8375 | 0.0168 | $0^{b}$ |
| CoRoT-7b | $8.5^{a}$ | 0.93 | 0.8536 | 0.0172 | $0^{b}$ |
| KOI-961b | 2.8604 | 0.13 | 1.2138 | 0.0116 | ?[b] |
| GJ 1214b | 6.357 | 0.153 | 1.58 | 0.014 | $< 0.27$ |
| Kepler-9d | 6.99 | 1.0 | 1.5928 | 0.0273 | ? [b] |
| KOI-961d | 0.9535 | 0.13 | 1.8516 | 0.0154 | ?[b] |
| GJ 876d | 6.67 | 0.334 | 1.9378 | 0.02 | $0.207 \pm 0.055^{b,c}$ |
| GJ 3634b | 7.0 | 0.45 | 2.6456 | 0.0287 | $0.08 \pm 0.057$ |
| Kepler-21b | $< 10.488$ | 1.34 | 2.7857 | 0.0425 | 0 |
| Kepler-18b | 6.897 | 0.972 | 3.5047 | 0.0447 | $0^{b}$ |
| Kepler-20b | 8.58141 | 0.912 | 3.6961 | 0.0454 | $< 0.320^{b}$ |
| CoRoT-7c | 8.39 | 0.93 | 3.698 | 0.046 | $0^{b}$ |
| HD 156668b | 4.16 | 0.772 | 4.646 | 0.05 | 0 |
| GJ 674b | 11.76 | 0.35 | 4.6938 | 0.039 | $0.2 \pm 0.02$ |
| HD 7924b | 9.217 | 0.832 | 5.3978 | 0.057 | $0.17 \pm 0.16$ |
| HD 45184b | 12.7132 | - | 5.8872 | 0.0638 | 0.3 |
| Kepler-20e | $< 3.083$ | 0.912 | 6.0985 | 0.0507 | ?[b] |
| GJ 176b | 8.42 | 0.49 | 8.7836 | 0.066 | 0 |
| HD 97658b | 6.356 | 0.85 | 9.4957 | 0.0797 | $0.13 \pm 0.06$ |
| HD 125595b | 14.30 | 0.76 | 9.67 | 0.078 | 0 |
| Kepler-11b | 4.3 | 0.95 | 10.3037 | 0.091 | 0 [b] |
| Kepler-11c | 13.507 | 0.95 | 13.025 | 0.106 | 0 [b] |
| Kepler-20f | 14.3 | 0.912 | 19.577 | 0.11 | ?[b] |
| Kepler-11d | 6.099 | 0.95 | 22.687 | 0.159 | 0 [b] |
| Kepler-11e | 8.40 | 0.95 | 31.9959 | 0.194 | 0 [b] |
| Kepler-11f | 2.3 | 0.95 | 46.6888 | 0.25 | 0 [b] |

When the rotation is neglected, the attained equilibrium figure is a Jeans spheroid with b = c. In this case, the value of $\epsilon$ is given by the above equation without the second term. A classical result shows that for a free synchronous rotating body, the polar deformation or oblateness (that is, $1 - \frac{c}{b}$) is three times smaller than the prolateness due to the companion (this holds when we neglect $\left(\frac{\Omega}{n}\right)^2$ in (3); see Danby 1988, p.121)[4].

The restriction b = c enables us to relate $\epsilon$ with the unknown quantity $\frac{B-A}{C}$. The moment of inertia of the spheroid can be written as

$$A = \frac{1}{5}m(b^2 + c^2), \qquad B = \frac{1}{5}m(a^2 + c^2), \qquad C = \frac{1}{5}m(a^2 + b^2). \tag{4}$$

---

[4] The dependence on $\left(\frac{\Omega}{n}\right)^2$ in (3) is in fact weak. For instance, in the case of 55 Cnc e, $\epsilon \sim 2.8 \times 10^{-2}$. If we neglect the rotation term and consider $\Omega = 0$, we have $\epsilon \sim 2.75 \times 10^{-2}$. These values are smaller than when the uncertainty of the radius of the planet is considered in the calculation of $\epsilon$. For instance, in the case of 55 Cnc e, Gillon et al. (2012) give $R = 2.17 \pm 0.10$ $R_E$, which results the maximum values: $\epsilon \sim 3.21 \times 10^{-2}$ ($\Omega = n$), $\epsilon \sim 3.14 \times 10^{-2}$ ($\Omega = 0$).

Hence, using the definition of $\epsilon$ and imposing b = c into the above equations, we have

$$\frac{A}{C} = \frac{2}{(1-\epsilon)^{-2} + 1} \simeq 1 - \epsilon - \frac{1}{2}\epsilon^2. \tag{5}$$

Therefore, at first order in $\epsilon$:

$$\frac{B-A}{C} \simeq \epsilon. \tag{6}$$

Note that, neglecting $\left(\frac{\Omega}{n}\right)^2$ and averaging the expression (3) for $\epsilon$ over an orbital period we obtain $\epsilon = \frac{15MR^3}{4ma^3}(1 - \frac{3}{2}e^2)$ plus higher order terms in powers of $e$. This result coincides with that found by Giampieri (2004) for $p = \Omega/n = 1$ and using $k_f = 1.5$, where $k_f$ is the fluid Love number, meaning that Equation (3) is only valid for the synchronous case. For comparison with Giampieri's results, one must relate the gravitational coefficient $C_{22}$ with $\epsilon$ by $\epsilon \sim \frac{B-A}{C} = 4C_{22}/\xi$, where $\xi = \frac{C}{mR^2}$ is the moment of inertia factor.

In order to test the validity of Equation (6) we estimate $\epsilon$ for some bodies of the Solar System. For Europa we estimate $\epsilon \sim 1.9 \times 10^{-3}$, a value for which the order of magnitude agrees with the current estimative of $\sim 1.5 \times 10^{-3}$. The latter value has been obtained from the numerical values of the gravitational coefficient $C_{22}$ and $\xi$ given in Van Hoolst et al. (2008). In the case of Titan, we obtain $C_{22} = 1.4 \times 10^{-5}$, in agreement with one reported value of about $1 \times 10^{-5}$ (Iess et al. 2010). For Io, we find $C_{22} = 3.4 \times 10^{-4}$, in concordance with Schubert et al. (2004), who reported $C_{22} = 5.6 \times 10^{-4}$.

The expression for $\epsilon$ provides the instantaneous equatorial prolateness due to the perturbation of the star and the planet rotation, in which the corresponding equilibrium figure is a Roche ellipsoid (see Ferraz-Mello et al. 2008). However, the expression for the equatorial deformation can be separated into static and periodic components (Rappaport et al. 1997; Giampieri 2004). On one hand, the static part is the one in which the body reacts as a liquid or perfect fluid to the stress. On the other hand, the periodic part arises when the body's orbit around its companion is non-circular. The deformations associated with these two components involve different times scales and are related to the fluid and dynamical Love numbers.

What is important in order to analyze the spin-orbit problem is the static (or permanent) component of the deformation. Indeed, the periodic components average to zero along an orbital period. Moreover, it is easy to see that only cases in which $p = 1/2, 1, 3/2, ...$, contribute to the static part (see Giampieri 2004). This means that one would expect a permanent deformation for some special spin-orbit configurations.

However, as shown in Giampieri (2004), the static deformation depends on the specific resonance in which the rotation is trapped. Moreover, the leading term of the prolateness is of the order of $e^{2|p-1|}$ (see Giampieri 2004, see also Correia and Rodríguez (2013)). Hence, a maximum deformation would occur for synchronous rotation. Indeed, most previous works have considered the special case of synchronous motion, mainly because this is the final configuration resulting from tidal evolution (see Correia 2009). However, when the possibility of non-synchronous resonances is taken into account, one must use the static deformation value corresponding to the specific trapping. For instance, equation (11) in Giampieri (2004) allows us estimate $\epsilon$ for the $3 : 2$ resonance:

$$\epsilon_{3:2} \sim \frac{B-A}{C} = \frac{7}{2}\frac{k_f}{\xi}\frac{m_0}{m}\left(\frac{R}{a}\right)^3 e, \tag{7}$$

where $e$ is the orbital eccentricity.

Finally we would like to note that, in the present study, we have assumed constant $\frac{B-A}{C}$ value. However, depending on the adjustment to hydrostatic equilibrium $\frac{B-A}{C}$ may itself be time-dependent. The time-scale for adjustment to the completely relaxed state depends on material parameters, e.g., rigidity and viscosity which are highly temperature dependent. Furthermore, it will depend on the spin-orbit coupling the planet is trapped in. For the application to terrestrial planets given here, the constant, i.e. frozen, $\frac{B-A}{C}$ configuration is justified because adjustment of solid rocky planets to equilibrium usually occurs on geological time-scales.

Therefore, the states described here do not necessarily represent hydrostatic configurations. For Jupiter-like planets, the simplification of constant $\frac{B-A}{C}$ values may not be justified because of the fluid behavior of the gas giants even in the short time-scales of their orbital periods. Adjustment to the hydrostatic state in a specific resonance will be relatively fast. In this case, the time-dependence of the gas giants $\frac{B-A}{C}$ must be taken into account in a coupled set of differential equations describing the spin, the orbit and the hydrostatic gravity field of the planet.

**Planet with unknown radius**

Table II has been calculated for planets with known estimated radii. However, for all other planets listed in Table I, $R$ is unknown. In these cases, we can estimate the order of magnitude of $\epsilon$ by analyzing its dependence with different parameters, as we illustrate below. Figures 1(a-e) show the level curves of the function $\epsilon = \epsilon(r, m)$ calculated using Equation (3) considering five fixed values of the radius of a fictitious planet of a star with mass $m_0 = 1 m_{SUN}$, where $m_{SUN}$ is the solar mass. The contours that appear in gray correspond to "small" value of $\epsilon < 0.01$, while the colored levels correspond to "large" prolateness where $\epsilon \geq 0.01$. We arbitrarily define here the value $\epsilon = 0.01$ as the limit between large and small prolateness of Super-Earths[5]. Figure 1(a) shows that exoplanets similar to the Earth (i.e., with mass and radius similar to the Earth's values) can have large equatorial prolateness for all $r < 0.018 AU$ (note the colored level curves near to the bottom left of the panel).

A zoom close to the origin of Figure 1(a) is shown in Figure 1(f). We can note that for small values of $(r, m)$ in the figure we have $\epsilon \sim 1$ (blue levels)[6].

Figure 1(f) suggests that a planet with radius $R \sim 1 R_E$, mass similar to the Earth's, and located in a very compact orbit ($r \ll 0.015$), can suffer important effects of the torque of the star on its rotation since its prolateness can be large (note in Figure 1(f) that the gray region is located above $m \sim 3.5 M_E$). The recently discovered KOI-961c, KOI-55b and KOI-55c are examples of this kind of planet (see Table I). As they are transiting planets, we can calculate their $\epsilon$ utilizing Equation (3). Table II shows that the prolateness of KOI-55b, KOI-55c are very large ($> 0.1$), which can be explained by their very small mass and small planet-star distance, in accordance with Figure 1(f) (however, Figure 1(f) refers to a star with $m_0 = 1 M_{SUN}$ while the mass of KOI-55 is half of this). In Section 3.1.2 we will study the dynamics of rotation of these planets with our model.

---

[5] As we will see in details in Section 3, for $\epsilon > 0.01$ the rotation may be highly disturbed by the torque of the central star in the case of eccentric orbits. However, this division between "small" and "large" prolateness is arbitrary and will be used in this study only for didactic purposes.

[6] We remark that very large values of $\epsilon$ close to the unit are not valid within the first-order approximation in $\epsilon$.

*Table II:* Numerical value of $\frac{B-A}{C} \simeq \epsilon$ for several Super-Earths with known radius. $\epsilon$ has been calculated with Equation (3) where we consider $\Omega = n$. The values of radius, in units of the Earth radius ($1R_E \simeq 6,378$ km), have been taken from http://exoplanet.eu/index.php.

| Planet | $\frac{B-A}{C}$ | $R\ (R_E)$ |
|---|---|---|
| KOI-55b | $2.60 \times 10^{-1}$ | $0.76^a$ |
| KOI-55c | $1.17 \times 10^{-1}$ | $0.87^a$ |
| 55 Cnc e | $2.80 \times 10^{-2}$ | $2.17^{+0.1}_{-0.1}$ |
| GJ 1214b | $1.77 \times 10^{-2}$ | $2.74^b$ |
| KOI-961c | $1.18 \times 10^{-2}$ | $0.73^{+0.2}_{-0.2}$ |
| Kepler-10b | $1.15 \times 10^{-2}$ | $1.42^{+0.033}_{-0.036}$ |
| CoRoT-7b | $9.92 \times 10^{-3}$ | $1.68^{+0.09}_{-0.09}$ |
| Kepler-9d | $3.04 \times 10^{-3}$ | $1.64^{+0.19}_{-0.14}$ |
| KOI-961b | $1.35 \times 10^{-3}$ | $0.78^{+0.22}_{-0.22}$ |
| Kepler-18b | $1.22 \times 10^{-3}$ | $2.00^{+0.1}_{-0.1}$ |
| Kepler-20b | $7.56 \times 10^{-4}$ | $1.90^{+0.12}_{-0.21}$ |
| Kepler-21b | $6.99 \times 10^{-4}$ | $1.63^{+0.04}_{-0.04}$ |
| KOI-961d | $6.70 \times 10^{-4}$ | $0.57^{+0.18}_{-0.18}$ |
| HD 97658b | $6.45 \times 10^{-4}$ | $2.93^{+0.28}_{-0.28}$ |
| Kepler-11b | $2.18 \times 10^{-4}$ | $1.97^{+0.19}_{-0.19}$ |
| Kepler-11c | $1.80 \times 10^{-4}$ | $3.15^{+0.3}_{-0.3}$ |
| Kepler-11d | $1.52 \times 10^{-4}$ | $3.43^{+0.32}_{-0.32}$ |
| Kepler-20e | $1.45 \times 10^{-4}$ | $0.868^{+0.074}_{-0.096}$ |
| Kepler-11e | $1.39 \times 10^{-4}$ | $4.52^{+0.43}_{-0.43}$ |
| Kepler-11f | $4.57 \times 10^{-5}$ | $2.61^{+0.25}_{-0.25}$ |
| Kepler-20f | $4.72 \times 10^{-6}$ | $1.00^{+0.1}_{-0.13}$ |

$^a$: Charpinet el al. (2011);

$^b$: The current estimative in http://exoplanet.eu/index.php is $2.84 \pm 0.2\ R_E$. GJ 1214b has a mass within the interval $m < 10M_E$, but as shown in Charbonneau et al. (2009) and Valencia (2011), due to its large radius, it may have a substantial gas layer and cannot be considered a solid-like planet.

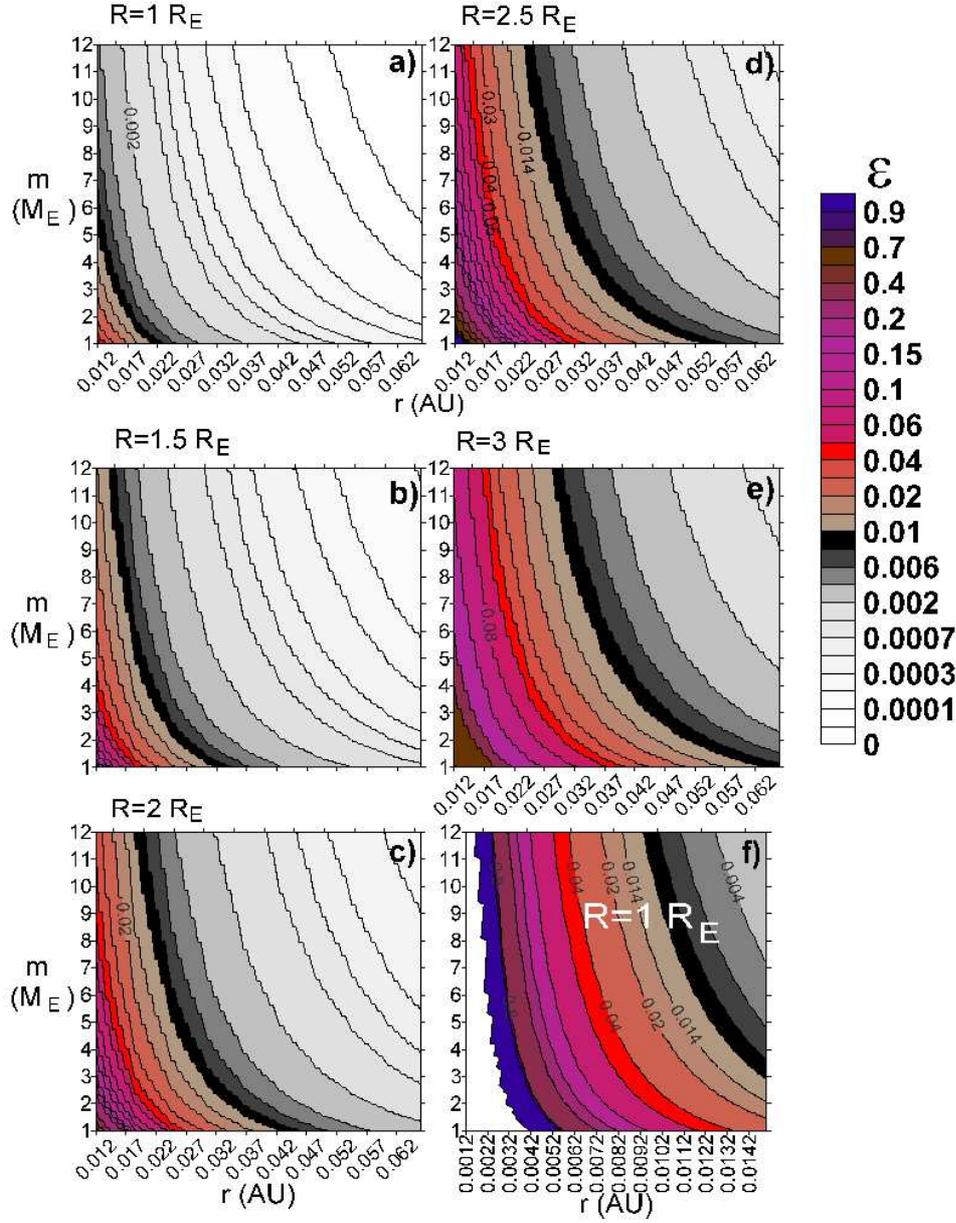

**Fig. 1** (a-e) Level curves of $\epsilon$ in the plane ($r \times m$) (star-planet distance ($r$) and the mass of the planet ($m$)), for different values of the radius of a fictitious planet orbiting a star with 1 $M_{SUN}$. (f) Zoom of Figure 1(a) close to the origin of the $r$-axis. The blue contour shown in (f) shows prohibitive levels with $\epsilon \sim 1$. The white portion of the figure close to the origin corresponds to level curves whose existence is not physically possible since they correspond to $\epsilon > 1$. The units are the Earth radius ($R_E$), the Earth mass ($M_E$), the solar mass ($M_{SUN}$) and the Astronomical Unit (AU).

## 3 Surfaces of Section and the Dynamics of Rotation

In this section we show surfaces of sections of the numerically computed solutions[7] of the model (1) for several Super-Earths listed in Table I. Following Wisdom et al. (1984), we investigate the dynamics of rotation in the plane $(\theta, \dot{\theta}/n)$ at $f = 0$, where $f$ is the true anomaly and $\dot{\theta}/n$ is the ratio of the angular rotational velocity and the mean motion. Thus, in the surfaces of sections, we plot the values of $(\theta, \dot{\theta}/n)$ at the moment of the passage of the planet through the pericenter in the case of eccentric orbit.

The surfaces of sections will be constructed for fixed values of $\frac{B-A}{C} \sim \epsilon$ in order to evaluate the stability of rotation of several planets. While the estimations of $\epsilon$ given by Equation (3) are only valid for synchronous motion, we can still recalculate the surfaces of section with the appropriate value of $\epsilon$ around each spin-orbit resonance (e.g. Equation (7) for the $3/2$ case). A maximum deformation is expected for synchronous rotation. Hence, when we use the Equation (3)) for the estimate of $\epsilon$, we would obtain an overestimation of the chaotic behavior resulting from the surfaces of section analysis. In Section 3.2.2 we will give an example in which we show the phase space of the spin-orbit dynamics adopting different $\epsilon$ for several resonances. In this case, the tool to analyze the stability will be the spectral analysis of the numerical solutions, since we cannot plot surfaces of sections around different resonances in the same figure when we adopt different $\epsilon$.

In order to present surfaces of sections, we consider four main types of Super-Earths: planets with circular or non-circular orbits, and those which are single planets or are present in *multiple* systems. We have organized the presentation of the results as follows. In Section 3.1, we review several topics of spin-orbit resonances and study examples of close-in planets in circular or quasi-circular orbits. In Section 3.2, we study the dynamics of rotation of close-in planets present in multiple systems considering eccentric orbits with orbital period $P < 7.5$ days[8]. In Section 3.3, we analyze the cases of single planets in both circular and eccentric orbits. In Section 3.4, we apply the Chirikov overlapping criterium to explore chaos in some systems around the domains of the 1:1 and 3:2 spin-orbit resonances.

### 3.1 Super-Earth in circular orbits

#### 3.1.1 CoRoT-7b, Kepler-10b

Figure 2 shows surfaces of sections for CoRoT-7b and Kepler-10b, two Super-Earths very close to their host stars ($a \sim 0.017$ AU). Figure 2(a) corresponds to the case of CoRoT-7b. Since its orbit is circular, the system (1), (2) is integrable in this case, and therefore we say that the dynamics of rotation of CoRoT-7b is regular, i.e., no chaotic motion is admissible in the domain of the model. We indicate in Figure 2(a) the domain of the synchronous motion located around the equilibrium point at $\theta = 0$, $\dot{\theta} = n$. The closed curves around the synchronous state show the amplitude of the physical libration of the major axis of the planet[9] relative to the star, whose motion is given quantitatively by the oscillation of the synodic angle $\psi = f - \theta$ around zero. In the section $\theta$ follows $\psi$ since we put $f = 0$.

In Figure 2(b) we consider a fictitious non-null value for the orbital eccentricity, $e = 0.01$. In this case, the domain of other spin-orbit resonances may appear in the sections, like the 3:2 resonance, for instance. In fact, it can be shown that the domains of low-order spin-orbit resonances exist in the phase space only in the case of $e \neq 0$. This can be understood as follows. The average, with respect to the mean-anomaly, of the torque of the star on the rotation of planets in circular orbits, is null for all low-order spin-orbit regimes except the synchronous one (Goldreich and Peale 1966). Thus, if the torque is null in a determined resonance, libration of the *corresponding synodic angle* is not possible. Therefore, in the case of circular orbits, this is a physical explanation for the absence of resonant islands in the surfaces of sections except those that are synchronous.

---

[7]  We use the code of Everhart (1986).

[8]  In the Section 4 we list the results for some planets with $7.5 < P < 33$ days.

[9]  Following classical astrodynamics texts (e.g. Danby 1988), we use the term *physical* librations to refer to motions associated with the non-spherical shape of the secondary body. The often called *optical* librations do not require an asymmetry of the secondary since they are not associated with a disturbed rotator. For instance, the optical libration in longitude exists due to differences between a (constant) velocity of rotation and the instantaneous orbital velocity in an eccentric orbit. In the disturbed (non-spherical) case, the optical libration is superposed with the forced component.

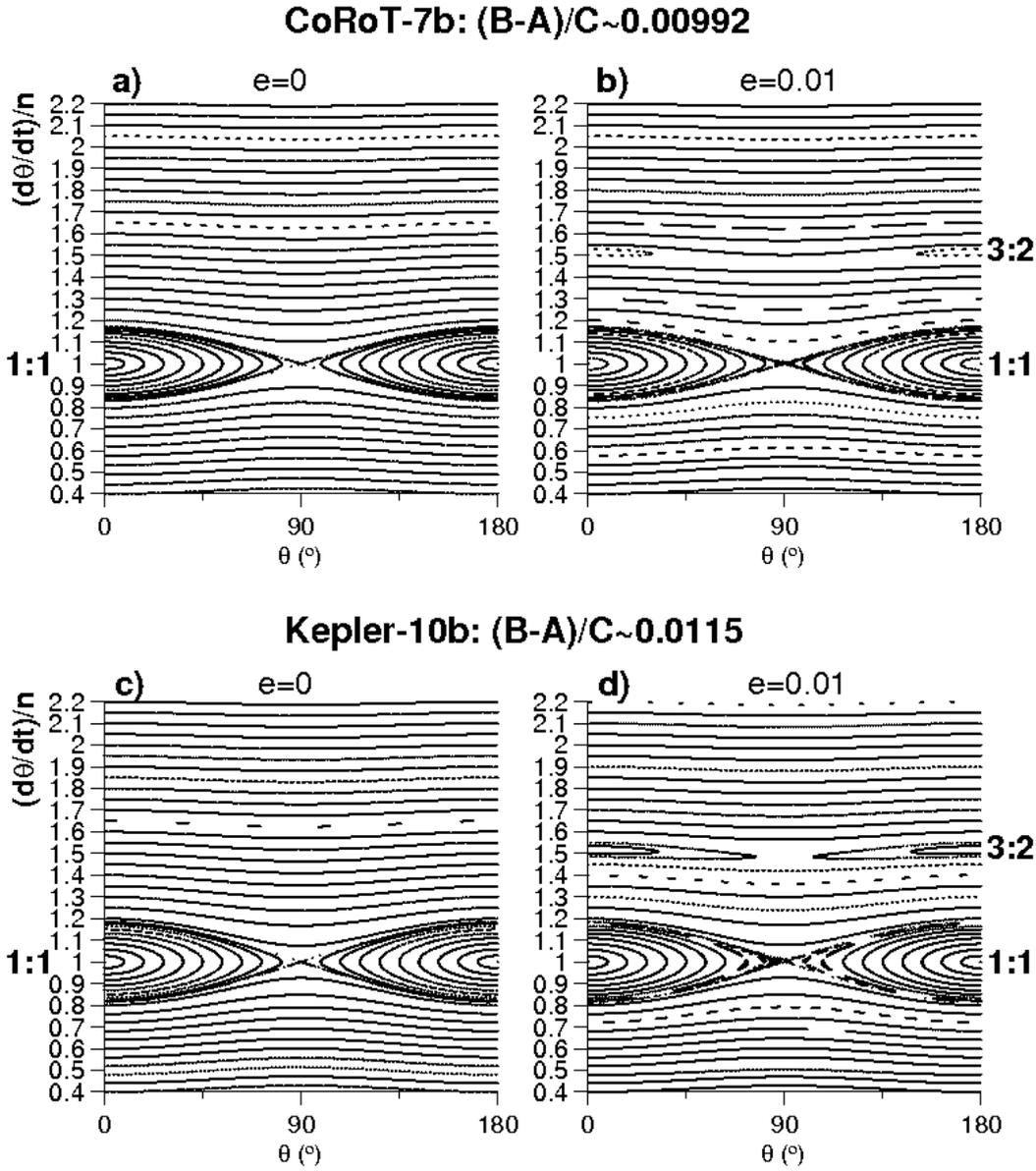

**Fig. 2** Surfaces of Section in the plane ($\dot{\theta}/n \times \theta$) of numerical solutions of (1). The sections have been made at each orbital revolution of the planets CoRoT-7b (Figures 2(a,b)) and Kepler-10b (Figures 2(c,d)). Each orbit has been integrated for 1140 orbital periods of the planet. The numerical value of $\frac{B-A}{C}$ has been calculated for the synchronous motion with Equation (3) and are given at the top of the panels (see also Table II). In the plots on the left, the orbital eccentricities correspond to the real values (i.e., the reported ones). In the plots on the right, the values of the eccentricities are fictitious and have been chosen in order to evaluate the domain of the 3:2 spin-orbit resonance in the case of quasi-circular orbit.

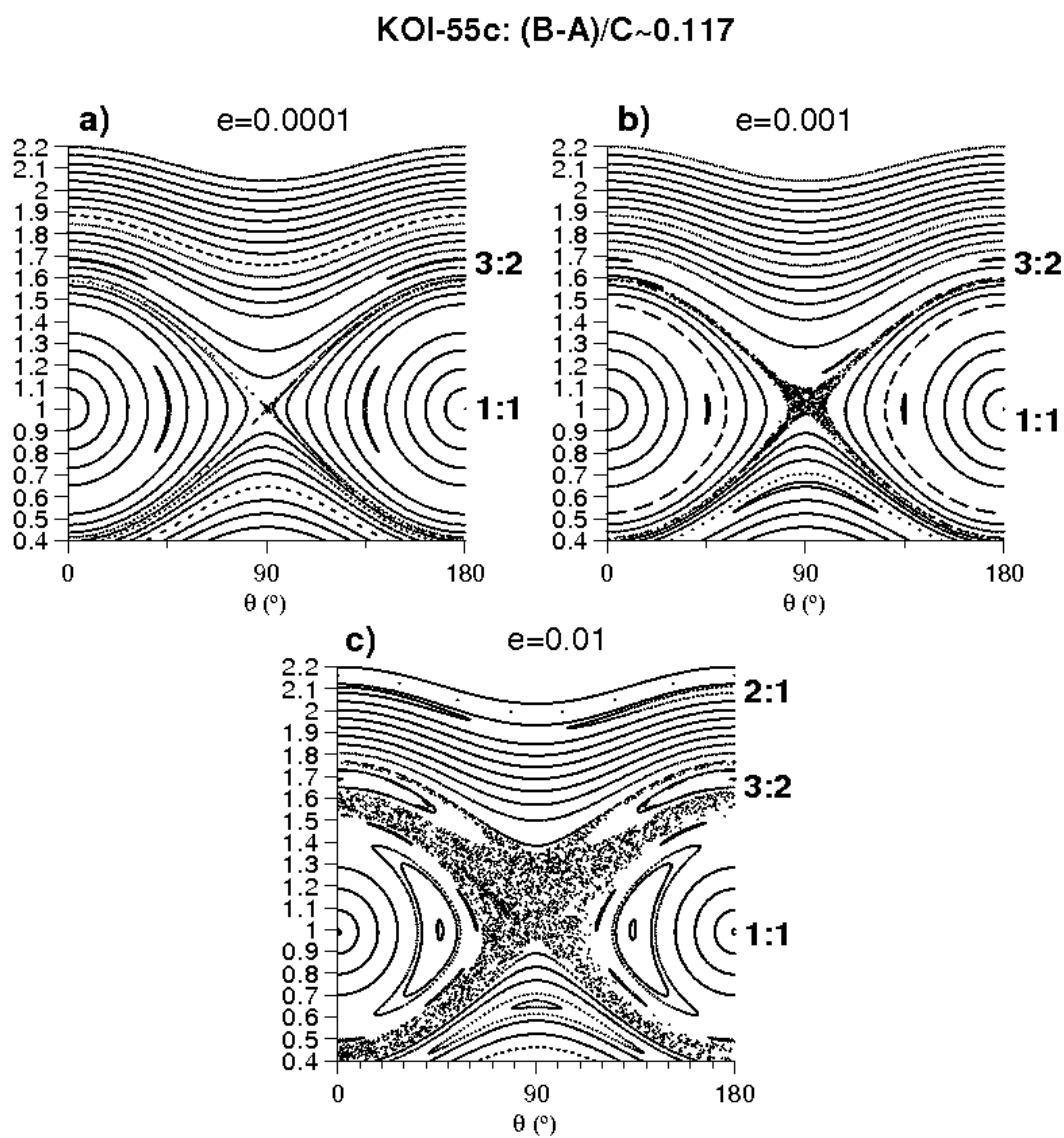

**Fig. 3** Surfaces of Section in the plane $(\dot{\theta}/n \times \theta)$ of numerical solutions of (1). The sections have been made at each orbital revolution of the planet and KOI-55c. Different fictitious values of the orbital eccentricity (indicated in each plot) have been considered. Each orbit has been integrated for 1140 orbital periods of the planet. The numerical value of $\frac{B-A}{C}$ has been calculated for the synchronous motion using Equation (3) and are presented at the top of the panels (see also Table II).

The domain of the synchronous motion is always present in the phase space even for $e = 0$. For $e = 0$ the average torque is never null except when $B - A = 0$ *or* is at the equilibrium point of the exact synchronism. For *circular orbits* the fixed point of the synchronous motion is located at $\dot{\theta} = n$ *and* $\theta = 0$. For $e \neq 0$, the fixed point corresponding to the equilibrium of the system is not located exactly at $\dot{\theta} = n$ and it is shifted compared to the case of null eccentricity. This can be seen by inspection of the fixed point of the 1:1 resonance in the surfaces of sections, mainly in the case of high values of eccentricity (e.g. Figures 5(d) and 6(c,d)). This shift occurs since the sections are done in the pericenter of the planet's orbit, where the instantaneous velocity is greater than the rotation rate. In the surface of section, we show that, at first order of eccentricity, the equilibrium point of the 1:1 resonance is located at $\frac{\dot{\theta}}{n} \simeq 1 - \kappa_{1:1} < 1$, where $\kappa_{1:1} \equiv \frac{2\omega_0^2 e}{n^2 - \omega_0^2}$, $\omega_0^2 \equiv 3n^2 \frac{B-A}{C}$ (see also Murray and Dermott 1999, pages 215-216).

For small eccentricity, it can be shown that physical libration of $\psi = \psi_{1:1} = f - \theta$ *includes* three components: forced and optical with frequency given by the mean-motion $n$, and long-term, free libration, with frequency $\omega_0$. The general expression of $\omega_0$ is given by $\omega_0^2 = 3n^2|H(p,e)|\frac{B-A}{C}$, where: $H(p,e)$ is the coefficient of the mean torque at the resonance of order $p$; $p$ is a half-integer which defines the resonance such that $p = 1$ for 1:1 resonance, $p = 2$ for 2:1, $p = +3/2$ for 3:2 etc (see Appendix B and appendix in Rodríguez et al. 2012)[10]. For 55 Cnc e, close to the 1:1 fixed point, the long-term period is about 2.54 days, while the forced component librates at each $\sim 0.736$ days. In the case of non-circular orbits, chaotic rotation is physically possible mainly close to the separatrices of the spin-orbit resonances. This regime of motion is possible since the physical libration can become unstable at greater amplitudes. In the example given in Figure 2(b), however, the values of eccentricity and $\frac{B-A}{C}$ are very small, and the chaotic layer is very thin.

Figures 2(c,d) show the case of Kepler-10 b. We can note the similarities with the system of CoRoT-7b discussed here.

### 3.1.2 KOI-55b, KOI-55c

In spite of regular motion analyzed in Figure 2, we can ask if a fictitious planet closer to its star than CoRoT-7b and Kepler 10-b would have similar properties of the rotation phase space even in the case of near-circular orbits. As discussed in Section 2, depending on the star-planet distance, a planet can have a very large value of $\epsilon \gg 0.01$. We give two examples of this kind of Earth-like candidate recently discovered: KOI-55b,c, located at 0.0060 AU (planet b) and 0.0076 AU (planet c) from the star, respectively (Table I). The values estimated for $\frac{B-A}{C}$ are $\sim 0.26$ (KOI-55b) and $\sim 0.117$ (KOI-55c) (Table II).

Figure 3 shows surfaces of section for KOI-55c. The numerical value of the orbital eccentricity is unknown (Charpinet el al. 2011). Due to its proximity to the star, it can be expected that its eccentricity is close to zero (see Section 1.4 and Appendix A). However, we consider in Figure 3 three different quasi-circular orbits. The domain of the 1:1 resonance is very large. The phase space shows the presence of other spin-orbit resonances, like the 3:2 resonance. The fixed points of all resonances except the synchronous one are slightly shifted from the exact commensurability. For instance, in Figure 4(a) we show the physical libration of $\psi_{3/2} = 2f - 3\theta$ adopting an initial condition within the 3:2 resonance indicated in Figure 3(b), where $\dot{\theta}_0 = 1.685n$ and not $\dot{\theta}_0 = 1.5n$ as would be expected[11].

---

[10] Note that $\omega_0$ is a linear function of $n$ with linear coefficient depending on the prolateness and $H(p,e)$. In the case of 1:1 resonance, at *first order* of eccentricity, the relation between $\omega_0$ and $n$ does not vary with the value of eccentricity since $H(1,e) = 1$. This does not occur for large eccentricity for other low-order resonances like 3:2, where $H(3/2,e) = \frac{7}{2}e$ at first order.

[11] Utilizing $\omega_0^2 = 3n^2|H(p,e)|\frac{B-A}{C}$ we can obtain the value 9.783 days for the period of free libration, which agrees with the period of the long-term oscillation given in the result of the numerical simulation of the full equations shown in Figure 4(a). Note also that $\frac{2\pi}{\omega_0} \sim 28.5\frac{2\pi}{n}$,

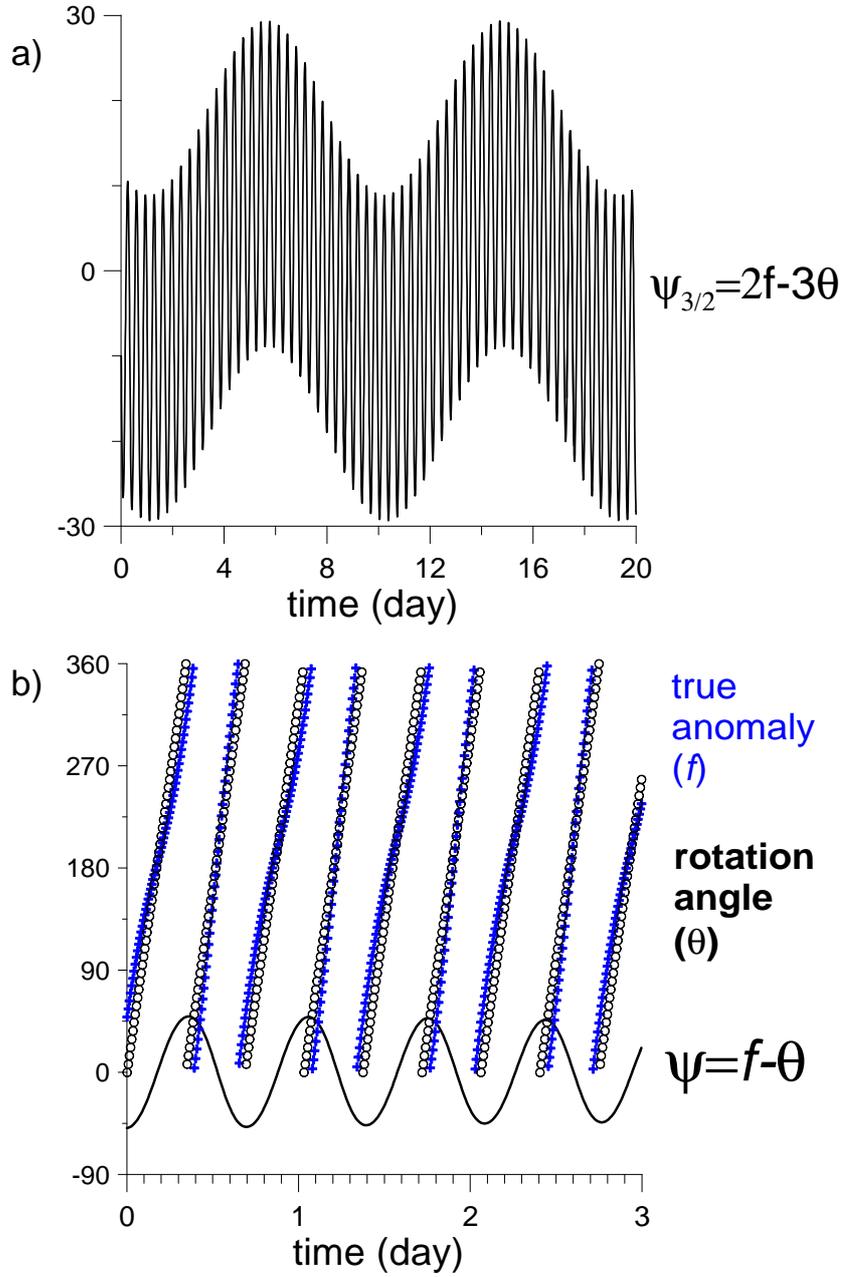

**Fig. 4** (a) Time evolution of $\psi_{3/2} = 2f - 3\theta$. The initial condition corresponds to the small island within the 3:2 resonance indicated in Figure 3(b): $\theta_0 = 5^o$, $\dot{\theta}_0 = 1.685n$, $f_0 = 0^o$, where $n$ is the mean motion. The long-term oscillation is the free libration while the short-term oscillation includes the forced and optical libration (see Section 3.1.1). (b) The time evolution of different quantities is indicated in the figure. The initial condition corresponds to the larger island of secondary resonance shown in Figure 3(c): $\theta_0 = 49^o$, $\dot{\theta}_0 = n$, $f_0 = 0^o$.

The phase space also shows secondary resonances within the synchronous island. Secondary resonances occur when the period of libration of the physical libration around the synchronous state is commensurate with the mean-motion of the planet. Figure 3(a) shows the 2:1 secondary resonance.

Figure 3(c) shows that setting $e = 0.01$, both, the chaotic layer around the separatrix of the 1:1 resonance, and the domain of 2:1 secondary resonance, have large domains in the phase space. An orbit close to the 2:1 secondary resonance is shown in Figure 4(b). Note that, while the true anomaly and the rotation angle evolve at approximately the same rate (period=0.34 days), the angle $\psi = f - \theta$ associated with the 1:1 resonance librates with twice the period of motion.

The dynamics of rotation of KOI-55b are very similar to the case of KOI-55c discussed above.

### 3.1.3 Other cases

The star KOI-961 hosts three transiting sub-Earth-sized planets (Muirhead et al. 2012). We estimate $\frac{B-A}{C}$ of these planets (Table II). KOI-961c has the larger value, $\epsilon \sim 0.018$; its dynamics of rotation are very similar to the cases presented above (e.g. Figure 2). The other two planets have relatively small values of $\epsilon \ll 0.01$ and the domain of the 1:1 resonance is very thin in their rotation phase space.

Earth-like candidate planets in *circular* orbits with semi-major axes larger than the previous cases have regular dynamics of rotation since their prolateness is small and the torque of the star is negligible. As examples we can cite the case of Kepler-21b, Kepler-18b, Kepler-20b and CoRoT-7c, close-in planets with orbital parameters with the same order of magnitude (see Table I). Kepler-21b, Kepler-18b and Kepler-20b are transiting planets and we can estimate their values of $\frac{B-A}{C}$.

Table II also includes the $\frac{B-A}{C}$ of Kepler-20e, Kepler-20f, and the close-in planets of the star Kepler-11.

## 3.2 Planets with eccentric orbits with $P < 7.5$ days present in multi-planetary systems

### 3.2.1 GJ 876d and Kepler-9d

Kepler-9d and GJ 876d are small planets belonging to multiple systems where the outer members are hot Jupiters and whose orbits are close to the 2/1 mean-motion resonances. They have similar orbital period and mass (see Table I), but orbit around stars with very different values of mass since GJ 876 is a M4 V star with mass $m_0 \sim 0.334 M_{SUN}$ (Correia et al. 2010) while Kepler-9 is a solar-type star with $m_0 \sim 1 M_{SUN}$ (Torres et al. 2011). Kepler-9d is a transiting planet whose radius is determined but its orbital eccentricity is not; on the other hand, the eccentricity of GJ 876d is known while its radius is not. Therefore we will consider the value of radius as a free parameter in the case of Gliese 876d, and the orbital eccentricity in the case of Kepler-9d.

If GJ 876d is a terrestrial-like, solid planet, Valencia et al. (2007b) show that its radius cannot be larger than $\sim 12,000$ km $\simeq 1.88 R_E$. Adopting this value, we calculate its prolateness: $\frac{B-A}{C} \sim \epsilon \sim 0.0036$. GJ 876d has one of the largest reported orbital eccentricities among all the close-in Earth-like candidates. In order to study its rotation, we use the values of the eccentricity given in Correia et al. (2010), $e = 0.124$, and $e = 0.207$ given in Rivera et al. (2010). The results are shown in Figures 5(a,b). Inspection of the surfaces of section shows that the rotation of Gliese 876d is probably regular even considering very eccentric orbits since $\epsilon$ is small.

The radius of GJ 876d is unknown, so we also study its dynamics with a larger value than that given above: we set $R \sim 3 R_E \simeq 19,000$ km, out of the range given by Valencia

---

and the factor 28.5 is comparable to the number of short-term oscillations seen over one period of the free libration.

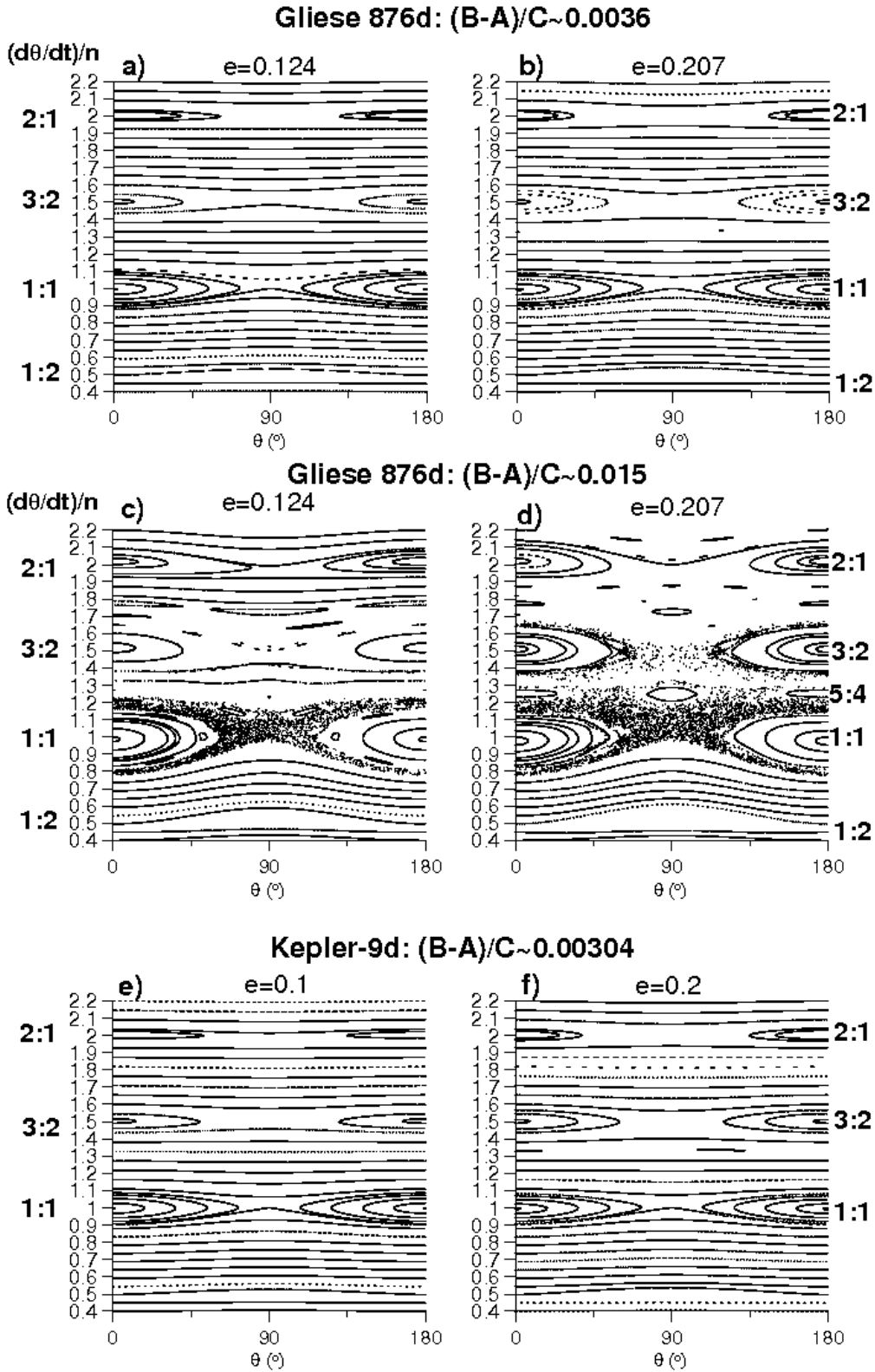

**Fig. 5** Surfaces of Section in the plane $(\dot{\theta}/n \times \theta)$ of numerical solutions of (1). The sections have been made at each orbital revolution of the planet Gliese 876d (figures (a-d)) and Kepler-9d (e,f). Each orbit has been integrated for 1140 orbital periods of the planet. $\frac{B-A}{C}$ for the synchronous motion of Kepler-9d is given in Table II; in the case of Gliese 876 d, the value of $\frac{B-A}{C}$ is discussed in Section 3.2.1. In (a,b) the orbital eccentricities correspond to two published values (see references in Section 3.2.1), while in (c,d) the eccentricities utilized in the simulations are fictitious.

et al. (2007b). For this radius, $\epsilon \sim 0.01$. Figures 5(c,d) show that the rotation may be more complicated in this case. For instance, we can see the overlap of the separatrix of the 3:2 and 1:1 resonances, engulfing the 5:4 resonance, when $e = 0.207$ (Figure 5(d))[12]. But it is important to note, however, that we have used the maximum value of $\epsilon$ in the whole phase space. As discussed in the second paragraph of Section 3, in such cases we are given an overestimation of chaotic regions of rotation motion around the resonances, which can in fact be smaller than those shown in Figures 5(c,d). Moreover, this case, where $R \sim 3R_E$ and $\epsilon > 0.01$, must be understood as a first view of a more complicated scenario since, in this case, GJ 876d cannot be considered a solid-like body (Valencia 2007b).

The eccentricity of Kepler-9d is unknown, but in analogy to the case of GJ 876d, we study the dynamics of its rotation in the case of eccentric orbit. We investigate the dynamics of Kepler-9d for two values of eccentricity: $e = 0.1$ (Figure 5(e)) and $e = 0.2$ (Figure 5(f)). In both cases, the rotation of Kepler-9d is regular in spite of eccentric orbits.

The similarities between Figures 5(a,b) and Figures 5(e,f) are evident. The value of $\epsilon$ estimated for Kepler-9d and GJ 876d used in these figures are similar: $\epsilon \sim 0.003$; we recall that the $\epsilon$ of GJ 876d was calculated assuming $R \simeq 1.88R_E$ given in the theory of Valencia et al. (2007b). Adopting the mass-radius relation of Valencia et al. (2007a), Kepler-9d is probably a solid-like body since $R \simeq 1.64R_E$ and our results on its dynamics of rotation are reliable in this case.

### 3.2.2 55 Cnc e

55 Cnc e is a low-mass planet orbiting a solar-like star (G8-K0) with very small period $< 1$ day (see Table I). With the recent estimate of its radius, probably this body is an Earth-like planet with important rock-component in its structure (e.g. Gillon et al. 2012)[13], a result which motivates us to study some aspects of its rotation with our model.

55 Cnc e has the second larger value of $\epsilon$ which we have calculated for all Super-Earths considered in this work (Table II). Figure 6(a-d) show four plots for this planet where we adopt different values for the orbital eccentricity. The dynamics of rotation is essentially regular in the immediate vicinity of the resonances for $e < 0.01$ (Figures 6(a,b)). For larger values, the results are shown in Figures 6(c,d). In the case corresponding to the larger published value of eccentricity ($e = 0.057$), there is a large chaotic region around the separatrix of the 1:1 resonance. For larger values of eccentricity, a large region of chaos located between the 1:1 and 3:2 appears in the phase space (Figure 6(d)). Note also the presence of 1:4 secondary resonance inside the synchronous island[14].

In order to confirm and to explore in greater detail the properties of the phase space given by the surfaces of section, we compute, in Figure 6(e), a dynamical map in a grid of initial conditions corresponding to the range of the axis given in Figure 6(c). In the dynamical maps, we plot the spectral number, $N$, defined by the number of peaks given in the numerical Fourier spectrum of some variable of the problem which are greater than a fraction of the highest amplitude (e.g. Callegari and Yokoyama 2010). We adopt a reference amplitude of 10% of the largest peak present in the spectrum of $\theta$. $N$ is represented on each point of the grid by a color, given in a gray scale, with the tonality dependent on $N$ in the following way: i) white points corresponding to $N = 1$; ii) gray points where $1 < N < N_c$, with $N_c$ being an arbitrarily large value of $N$; iii) black points with $N \geq N_c$. We adopt $N_c = 100$ in Figures 6(e,f). Distinct coloured regions of the grid suggest different regions of stability of the phase space: in white and gray regions, the motion is regular, where the spectra of the solutions contains a small number of separated and countable peaks. In darker regions, where the spectrum contains a large number of peaks, the motion can be chaotic, and in general the peaks are not well separated in the spectrum. Comparison and inspection of Figures 6(c,e) show us that the dynamical map agrees very well with the main dynamical regions of the system obtained with the surfaces of section.

---

[12] Our numerical experiments show that in the cases of strong perturbation, the domain of the 1:2 resonance is preserved from the chaotic regions located close to it. Wisdom (1984), in their study on the rotation of the satellite Hyperion, discuss this property of the 1:2 resonance. In fact, the averaged torques associated with the 1:1 and 1:2 spin-orbit resonances show that they are the strongest ones among all low-order resonances (e.g. Goldreich and Peale 1966).

[13] According to Valencia (2007a), supposing a solid-like structure for 55 Cnc e, its radius cannot be larger than $\simeq 1.8R_E$. Since the estimated radius of 55 Cnc e is $\simeq 2.17R_E$ it cannot be considered a pure solid-like planet, admitting a non-negligible fraction of other non-solid components in its composition. However, recent calculations with models of interior structure of 55 Cnc e show the possibility of its solid-like nature (Madhusudhan et al. 2012). Moreover, aiming to improve our model, it could be interesting to consider the perturbations in the rotation of the planet adopting different layers (e.g. van Hoolst et al. 2008).

[14] Wisdom (2004) shows that, in the case of regular satellites of the Solar System, a capture in such rotational state may enhance the tidal heating in the satellite interior by several orders of magnitude. The same would occur for close-in Super-Earths.

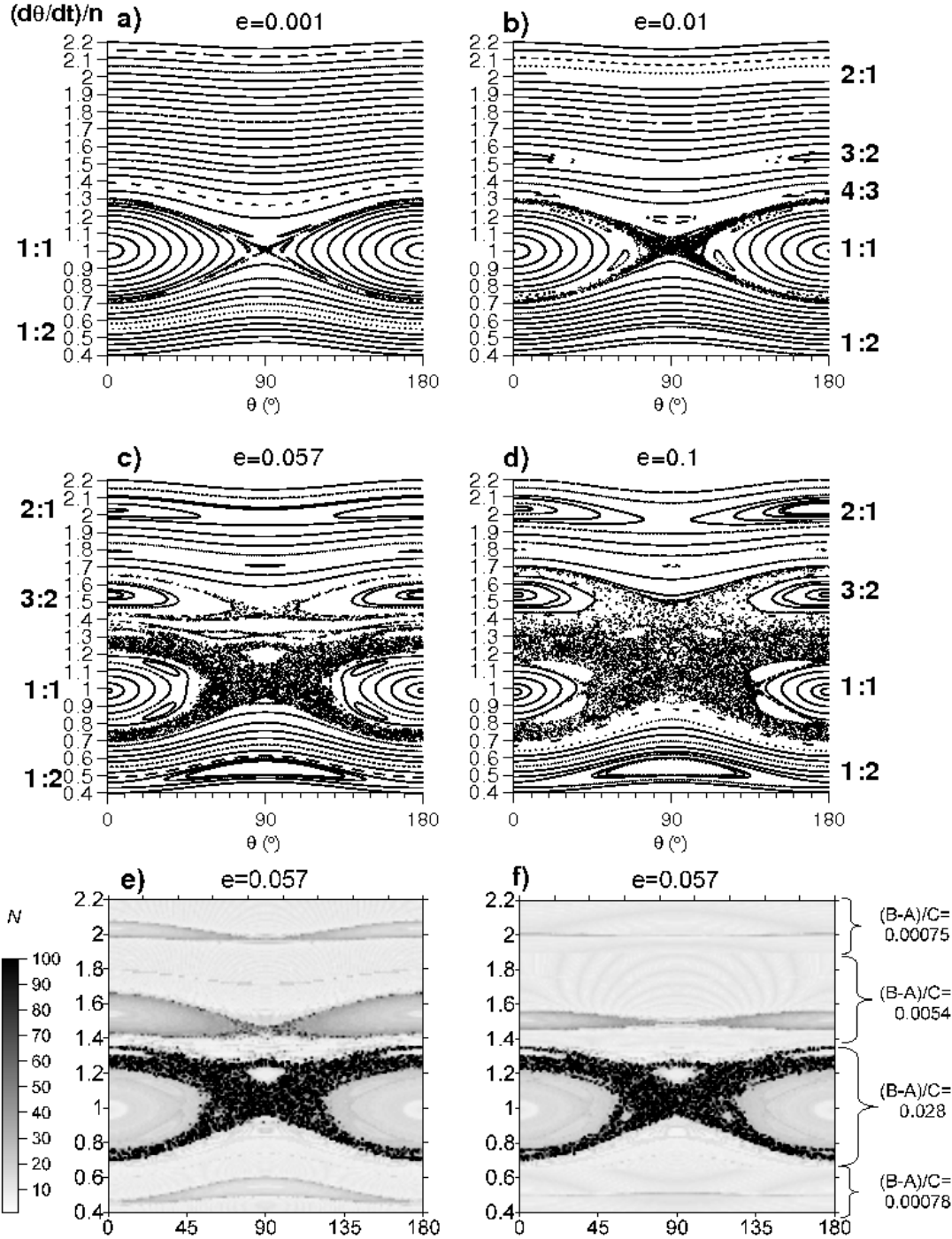

**Fig. 6** Surfaces of Section in the plane ($\dot{\theta}/n \times \theta$) of numerical solutions of (1). The sections have been made at each orbital revolution of the planet 55 Cancri and adopting different values of the orbital eccentricity which are indicated at the top of the plots. Each orbit has been integrated for 1140 orbital periods of the planet. The current (published) value of eccentricity is considered in (c) while the other values are fictitious. (e) Dynamical map (see Section 3.2.2) in a grid of 22,801 points corresponding to (c). In (a-e): $\frac{B-A}{C} \sim \epsilon = 0.028$ for the synchronous motion is estimated adopting $R = 2.17 R_E$ for the radius of the planet. (f) The same as (e) but adopting different values of $\frac{B-A}{C}$ around different resonances (see Section 3.2.2).

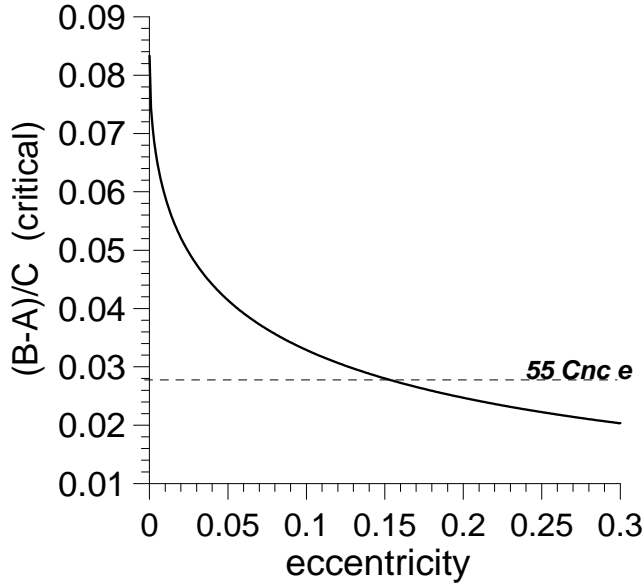

**Fig. 7** Critical values of $\frac{B-A}{C}$ for overlap of the 1:1 and 3:2 spin-orbit resonances calculated from Equation (8), Section 3.4. The horizontal dashed lines indicate the values of $\frac{B-A}{C}$ of the planets indicated in the figure.

In order to investigate the phase space of rotation of 55 Cnc e adopting different values of prolateness around distinct resonances, we recalculate Figure 6(e) with the values of $\frac{B-A}{C} \sim \epsilon$ listed in Figure 6(f). The value of $\frac{B-A}{C}$ in the vicinity of the 3:2 spin-orbit resonance has been calculated with Equation (7) where we put $\xi = 0.4$ and $k_f = 1.5$, which constitute values for homogeneous spherical bodies (see Ferraz-Mello et al. 2008). For 2:1 and 1:2 resonances, similar expressions can be found, building on Giampieri's (2004) work. Figure 6(f) shows that the domains of all these resonances except 1:1 are smaller in the phase space when compared to the case where the largest value of $\frac{B-A}{C} \sim 0.028$ is (i.e., the value estimated for the synchronous motion).

We conclude that a non-null value of the eccentricity of 55 Cnc e, and its proximity to the star, lead to a strong periodic perturbation on its rotation. Thus, according to our model, in an evolutionary scenario where some dissipative mechanism is able to reduce the primordial spin of the planet to the synchronous resonance, 55 Cnc e would have crossed a chaotic layer before synchronism was attained, similar to the history of the rotation dynamics of the irregularly shaped, close-in, satellites of the solar system (Wisdom et al. 1984, Wisdom 1987, Wisdom 2004). We note, however, that, as shown in Rodríguez et al. (2012) in their scenario of tidal evolution, the rotation of 55 Cnc e can evolve to synchronism through successive temporary trappings in spin-orbit resonances.

### 3.3 Systems with a single planet

We consider in this separated section some aspects of the dynamics of rotation of stars with a single planet in both, circular and eccentric orbits.

The systems with single planets in *circular* orbit which we study are: HD 156668b (orbital period $P = 4.646 days$, $m = 4.16 M_E$), GJ 176b ($P = 8.78 days$, $m = 8.42 M_E$) and HD 125595b ($P = 9.67 days$, $m = 14.3 M_E$). The radii of all of them are unknown quantities, but we can use figures similar to Figure 1, Section 2, to infer that their prolateness is negligible due to their relatively large mass and the range of planet-star distance $a > 0.05 AU$. Therefore their rotations suffer very small perturbation from their respective stars.

There are several examples of isolated Super-Earths with reported *eccentric* orbits: GJ 3634b ($P = 2.64 days$, $e = 0.08$, $m = 7.0 M_E$), GJ 674b ($P = 4.7 days$, $e = 0.1$, $m = 11.76 M_E$), HD 7924b ($P = 5.4 days$, $e = 0.17$, $m = 9.27 M_E$), HD 45184b ($P = 5.88 days$, $e = 0.3$, $m = 12.7 M_E$), HD 97658b ($P = 9.5 days$, $e = 0.17$, $m = 8.3 M_E$). Except in the case of GJ 3634b, we can note that their masses are large and their radius would also be large in the scenario of a solid-like planet presented in Valencia et al. (2007a). However in all cases, the prolateness is $\epsilon \ll 0.01$ and the dynamics of rotation are regular, like several other cases shown previously, in spite of their eccentric orbit.

The case of GJ 3634b is different due to its proximity to GJ 3634 and relatively small mass. In this case, due to the similarities of the values of the parameters of the star-planet system with the system GJ 876d, we can use the results of the Section 3.2.1 to infer that the dynamics of rotation is regular in spite of a probably non-negligible value of prolateness.

### 3.4 The Chirikov's overlap criterium

We can use Chirikov's overlap criterium to estimate local chaos around the main resonances in terms of the main parameters of the problem: $\frac{B-A}{C}$ and eccentricity. In fact, the separatrices of 1:1 and 3:2 resonances overlap for all values of $\frac{B-A}{C}$ larger than a critical quantity which is given by the following expression:

$$\left(\frac{B-A}{C}\right)_{critical} \sim \frac{1}{3}\left(\frac{1}{2+\sqrt{14e}}\right)^2. \qquad (8)$$

(Wisdom et al. 1984).

Figure 7 shows the plot of the critical $\left(\frac{B-A}{C}\right)_{critical}$ as a function of eccentricity. We indicate the value of $\epsilon = 0.028$ calculated for 55 Cnc e, and for $e = 0.15$, the overlap occurs. Surfaces of section show good agreement with Chirikov's overlap criterium: for $e < 0.1$ the separatrices of the 1:1 and 3:2 do not overlap (see Figures 6(b,c)), but for $e = 0.1$ (Figure 6(d)), a value very close to the critical one given in Figure 7, the overlap occurs.

Figure 7 shows that for large values of $\frac{B-A}{C} \sim 10^{-1}$ the overlap may occur for very small values of eccentricity $\sim < 0.01$. This is shown in Figure 3(c) in the case of KOI-55c.

## 4 Conclusions and Discussion

We have explored the dynamics of rotation of close-in exoplanets with a model often applied in studies of physical libration of natural satellites and Mercury (Goldreich and Peale 1966, Henrard 1985, Bills et al. 2009). This model is suitable to quantify the main effects of the gravitational torque of the star on the rotation of a planet when the obliquity of the axis of rotation of the planet is considered fixed (i.e., we neglect the motion in attitude).

The model is valid in the hypothesis of a rigid planet, and therefore we neglect in our calculations the effects of tidal torques of the central star on the rotation of the planet. Thus, we study the conservative dynamics of rotation of several possible Earth-like exoplanets. This type of planet, with a mass of $m < 10 M_E$, may have a permanent solid structure depending also on the value of the radius of the planet (e.g. Valencia et al. 2007a, 2011). Our methodology was based on previous studies like Wisdom et al. (1984, 1987, 2004), where the dynamics of rotation of regular satellites of Saturn, Phobos and Deimos are numerically explored through analysis of surfaces of section.

According *to the conservative model*, the effects of the torque of the star on the rotation of the planet depend basically on two parameters: the eccentricity of the orbit, and the existence of a permanent equatorial bulge (the prolateness, $\epsilon$) of an ellipsoid with semi-axes a > b > c. A non-null value of the prolateness implies that $\frac{B-A}{C} \neq 0$, where $A, B, C$ are the main moments of inertia which, in the case of exoplanets, are entirely unknown quantities. We show in this study that $\frac{B-A}{C} \sim \epsilon$, a relation valid at first order in $\epsilon$ when b = c. On the other hand $\epsilon$ can be written as a function of the masses of the system, the star-planet distance and the radius of the planet, and when these parameters are known, we can estimate an order of magnitude of $\epsilon$. This estimate of $\epsilon$ depends on the resonance we are considering. In general the transiting planets have a determination of the radius what allows us to estimate $\epsilon$. When the planet is not a transiting one, we show here how to obtain a significant range of the planet's prolateness in terms of the various parameters of the star-planet system.

We discuss the cases of Super-Earth candidates immersed in multi-planetary systems and those which are single planets. We also consider planets in circular and eccentric orbits. In the case of *circular* orbits the dynamics of rotation of exoplanets is trivial, i.e., it is regular and the synchronous state governs the phase space with a domain dictated only by the magnitude of the prolateness of the equator and the mass of the central star. The most interesting cases are close-in exoplanets with eccentric orbits, since in this case, the dynamics of physical libration may deviate significantly from the pendulum analogy. Large-scale chaotic motion of rotation for initial conditions close to the separatrix of the main resonances appears due to a combination of large prolateness and high eccentricity.

Our main results regarding the dynamics of rotation of some cases presented here are summarized below:

• KOI-55b,c, close-in planets with very small mass ($m < 1M_E$) have large prolateness $0.1 < \epsilon < 0.2$ and, even for small eccentricity ($e \sim 0.007$), there is the possibility of a complicated dynamics of rotation (i.e., chaotic motion, secondary resonance) of these planets around the synchronous motion.

• The rotation of planets like Kepler-10b, CoRoT-7b, KOI-961c can suffer strong effects of the star due to the relatively large values of their prolateness, which are given by the interval: $0.01 < \epsilon < 0.02$. However, their orbits are probably circular and the regions of rotation phase space are regular and dominated by the synchronous island, which is probably the loci of the current state of the rotations of these planets. The same occurs for other Super-Earths like CoRoT-7c and KOI-961b which however have small prolateness $\epsilon \ll 0.01$ and quasi-circular orbits, and the phase space is dominated by circulation regimes of motion where the width of the synchronous resonance is very small in the phase space.

• Close-in bodies in *eccentric* orbits and orbital period $P < 7.5$ days can have complicated dynamics depending mainly on the values of the radius ($R$) and mass ($m$) of the planet. For instance, GJ 876d can have large prolateness $\epsilon > 0.01$ if $R \sim 3R_E$. However, Kepler-9d, with mass similar to the GJ 876d's and known radius $\sim 1.64R_E$, presents regular motion of rotation even for eccentricities larger than 0.2.

• 55 Cnc-e admits large chaotic regions close the synchronous rotation adopting its current parameters ($e \sim 0.057$, $m \sim 8.58M_E$, $R = 2.17R_E$, $\epsilon \sim 0.028$). The implications of the chaos on evolution through spin-orbit resonance have been investigated in Rodríguez et al. (2012). The above value of prolateness has been estimated for synchronous resonance. However, as we show in this paper, based in previous works (e.g. Giampieri 2004), different resonances lead to distinct figures of equilibrium of the Super-Earths. A project to investigate the consequences of evolution through resonances by adopting different $\epsilon$, similar to the work of Rodríguez et al. (2012) is currently underway .

• The dynamics of planets *with eccentric orbits* and ($P > 7.5$ days) were also studied here (e.g. Gl 581c, HD 181433b, HD 215457c). We have shown that all of them have negligible prolateness when we assume a solid structure for them. Therefore they probably have regular motion of rotation around the synchronous or other main spin-orbit resonances.

• The prolateness in the synchronous resonance of several single Earth-like candidate planets considered here (e.g. GJ 176b, GJ 674b, HD 97658b) may be negligible due to their large values of mass and distance from the star. Therefore, their dynamics of rotation are regular in spite of eccentric orbits (GJ 674b, HD 97658b). On the other hand the planet GJ 3634b has a (maximum) probably non-negligible prolateness $\epsilon \sim 10^{-3}$. With this value as reference, Rodríguez et al. (2012) show simulations where this planet may be currently captured into the 2:1 spin-orbit resonance.

All simulations show that the dynamics of rotation of close-in *eccentric* Earth-like planets may have a complex behavior close to the regions of separatrices in the phase space depending on the prolateness. In a scenario including the action of a dissipative force which acts to spin down the primordial rotation, one should understand how planetary rotation of all bodies studied here evolves to synchronism traversing a chaotic region of the phase space (e.g. Rodríguez et al. 2012)[15]. Even in those cases with almost circular orbits, the rotation can be driven to a complex motion when $\frac{B-A}{C}$ is large (e.g. KOI-55b,c).

## Discussion on Tidal Evolution

● Due to tidal effects it is expected that the orbit of a close-in planet be circular. In Appendix A we calculate the timescale for circularization of a planet with elements similar to CoRoT-7b. We confirm previous results which show that planets with orbital period $P > 8$ days may have a non-circular orbit since the time scale for circularization is very large: $> 10$ billion years. Since the value of orbital eccentricity of some planets is still being improved (see Tadeu dos Santos et al. 2012) we consider in our calculations several values of eccentricity in addition to the ones given in the literature.

● We can also apply our calculations given in Appendix A for several single planets discussed in Section 3.4. The orbital periods of GJ 3634b, GJ 674b, HD 7924b and HD 45184b are shorter than 6 days, and thus a circular orbit is expected in light of the tidal effect. On the other hand, a null eccentricity is not expected for HD 125595b since $P \sim 9.4$ days.

● Although we have neglected the effects of tidal torques in this study, they would be important in the case of a non-rigid planets located very close to the central star. In Appendix B, we study the case of the planet CoRoT-7b utilizing a model of tides where the phase lags vary linearly with the corresponding tidal frequency. We calculate the magnitude of the tidal torque and compare it with the gravitational torque due to the equatorial prolateness in order to evaluate the stability of two spin-orbit resonances, the synchronous and 3:2 spin-orbit resonances. We found that the 1:1 resonance is stable adopting our estimative of $\epsilon = 0.0092$ for CoRoT-7b.

**Acknowledgements**  FAPESP (2006/58000-2 (NCJ); 2009/16900-5 (ARC).)

## References

1. Anglada-Escudé, G.; Lpez-Morales, M.; Chambers, John E.. How Eccentric Orbital Solutions Can Hide Planetary Systems in 2:1 Resonant Orbits. The Astrophysical Journal, **709**, 168-178 (2010).
2. Baraffe, I.; Chabrier, G.; Barman, T.. The physical properties of extra-solar planets. Reports on Progress in Physics, **73**, 1, pp. 016901 (2010).
3. Barnes, Jason W.; Fortney, Jonathan J.. Measuring the Oblateness and Rotation of Transiting Extrasolar Giant Planets. The Astrophysical Journal, **588**, 545-556 (2003).
4. Batalha, Natalie M.; Borucki, William J.; Bryson, Stephen T.; Buchhave, Lars A.; Caldwell, Douglas A.; Christensen-Dalsgaard, Jrgen; Ciardi, David; Dunham, Edward W.; Fressin, Francois; Gautier, Thomas N., III; and 42 coauthors. Kepler's First Rocky Planet: Kepler-10b. The Astrophysical Journal, **729** (2011).
5. Batygin, K.; Bodenheimer, P.; Laughlin, G.. Determination of the Interior Structure of Transiting Planets in Multiple-Planet Systems. The Astrophysical Journal Letters, **704**, L49-L53 (2009).
6. Beutler, G.. Methods of Celestial Mechanics, Vol. I (Springer, Berlin) (2005).
7. Bills, B. G.; Nimmo, F.; Karatekin, O.; van Hoolst, T.; Rambaux, N.; Levrard, B.; Laskar, J.. Rotational Dynamics of Europa. In: Europa, Edited by Robert T. Pappalardo, William B. McKinnon, Krishan K. Khurana; with the assistance of René Dotson with 85 collaborating authors. University of Arizona Press, Tucson. The University of Arizona space science series ISBN: 9780816528448, p.119 (2009).

---

[15]  Moreover, evolution through resonances in the case of strong perturbation can result in the chaotic tumbling of the rotation axis of the planets, similarly to the case of some Solar System's bodies (Wisdom et al. 1984; 1987).

8. Callegari Jr., N., Yokoyama, T.. Numerical exploration of resonant dynamics in the system of Saturnian inner Satellites. Planetary and Space Science, **58**, 1906-1921 (2010).

9. Carter, Joshua A.; Winn, Joshua N.. Empirical Constraints on the Oblateness of an Exoplanet. The Astrophysical Journal, **709**, 1219-1229 (2010).

10. Castan, T., Menou, K.. Atmospheres of Hot Super-Earths. The Astrophysical Journal Letters, **743**, Issue 2, article id. L36 (2011).

11. Celletti, A.; Voyatzis, G.. Regions of stability in rotational dynamics. Celestial Mechanics and Dynamical Astronomy, **107**, 101-113 (2010).

12. Charbonneau, David; Berta, Zachory K.; Irwin, Jonathan; Burke, Christopher J.; Nutzman, Philip; Buchhave, Lars A.; Lovis, Christophe; Bonfils, Xavier; Latham, David W.; Udry, Stphane; and 9 coauthors. A super-Earth transiting a nearby low-mass star. Nature, **462**, 891-894 (2009).

13. Charpinet, S.; Fontaine, G.; Brassard, P.; Green, E. M.; van Grootel, V.; Randall, S. K.; Silvotti, R.; Baran, A. S.; stensen, R. H.; Kawaler, S. D.; Telting, J. H.. A compact system of small planets around a former red-giant star. Nature, **480**, 496-499 (2011).

14. Correia, A. C. M., Levrard, B., Laskar, J.. On the equilibrium rotation of Earth-like extrasolar planets. Astronomy and Astrophysics, **488**, L63-L66 (2008).

15. Correia, A. C. M.. Secular Evolution of a Satellite by Tidal Effect: Application to Triton. The Astrophysical Journal Letters, **704**, L1-L4 (2009).

16. Correia, A. C. M.; Couetdic, J.; Laskar, J.; Bonfils, X.; Mayor, M.; Bertaux, J.-L.; Bouchy, F.; Delfosse, X.; Forveille, T.; Lovis, C.; Pepe, F.; Perrier, C.; Queloz, D.; Udry, S.. The HARPS search for southern extra-solar planets. XIX. Characterization and dynamics of the GJ 876 planetary system. Astronomy and Astrophysics, **511**, id.A21 (2010).

17. Correia, A. C. M.; Bou, Gwenal; Laskar, Jacques. Pumping the Eccentricity of Exoplanets by Tidal Effect. The Astrophysical Journal Letters, **744**, Issue 2, article id. L23 (2012).

18. Correia, A. C. M., Rodríguez, A.. On the equilibrium figure of close-in planets and satellites. The Astrophysical Journal, 767:128 (5pp) (2013).

19. Danby, J. M. A.. Fundamentals of Celestial Mechanics, 2nd edition. Willmann-Bell Richmond (1988).

20. Dobrovolskis, A. R.. Spin states and climates of eccentric exoplanets. Icarus, **192**, 1-23 (2007).

21. Dobrovolskis, A. R.. Insolation patterns on synchronous exoplanets with obliquity. Icarus, **2004**, 1-10 (2009).

22. Dobbs-Dixon, I, Lin, D. N. C., Mardling, R. A.. Spin-Orbit Evolution of Short-Period Planets. The Astrophysical Journal, **610**, 464-476 (2004).

23. Everhart, E.. An efficient integrator that uses Gauss-Radau spacings. In: IAU Coloquium **83**, 185-202 (1985).

24. Ferraz-Mello, S., Rodríguez, A., Hussmann, H.. Tidal friction in close-in satellites and exoplanets: The Darwin theory re-visited. Celest. Mech. Dyn. Astr., **101**, 171-201 (2008).

25. Ferraz-Mello, S.; Tadeu dos Santos, M.; Beauge, C.; Michtchenko, T. A.; Rodríguez, A.. On planetary mass determination in the case of super-Earths orbiting active stars. The case of the CoRoT-7 system. A&A, 531A (2011).

26. Giampieri, Giacomo. A note on the tidally induced potential of a satellite in eccentric orbit. Icarus, **167**, 228-230 (2004).

27. Gillon, M.; Demory, B.-O.; Benneke, B.; Valencia, D.; Deming, D.; Seager, S.; Lovis, Ch.; Mayor, M.; Pepe, F.; Queloz, D.; Sgransan, D.; Udry, S.. Astronomy & Astrophysics, Volume 539, id.A28 eprint (2012).

28. Goldreich, P., Peale, S.. Spin-orbit coupling in the solar system. The Astronomical Journal, **71**, 425-437 (1966).

29. Henrard, J. Spin-Orbit Resonance and the Adiabatic Invarint. In: S. Ferraz-Mello and W. Sessin (eds), Resonances in the Motion of the Planets, Satellites and Asteroids, IAG/USP, Sao Paulo, 19-26 (1985).

30. Hébrard, G.; Ehrenreich, D.; Bouchy, F.; Delfosse, X.; Moutou, C.; Arnold, L.; Boisse, I.; Bonfils, X.; Daz, R. F.; Eggenberger, A.; Forveille, T.; Lagrange, A.-M.; Lovis, C.; Pepe, F.; Perrier, C.; Queloz, D.; Santerne, A.; Santos, N. C.; Sgransan, D.; Udry, S.; Vidal-Madjar, A.. The retrograde orbit of the HAT-P-6b exoplanet. Astronomy & Astrophysics, **527**, id. L11 (2011).

31. Iess L., Rappaport N. J., Jacobson R. A., et al.. Science, **327**, 1367 (2010).

32. Jackson, B., Greenberg, R., Barnes, R.. Tidal Evolution of Close-in Extrasolar Planets. The Astrophysical Journal, **678**, 1396-1406 (2008a).

33. Jackson, B., Barnes, R., Greenberg, R.. Tidal heating of terrestrial extrasolar planets and implications for their habitability. Mon. Not. R. Astron. Soc., **391**, 237-245 (2008b).

34. Jackson, B., Greenberg, R., Barnes, R.. Tidal heating of Extrasolar Planets. The Astrophysical Journal, **681**, 1631-1638 (2008c).

35. Kitiashvili, I. N.; ALexander, G.. Rotational evolution of exoplanets under the action of gravitational and magnetic perturbations. Celestial Mechanics and Dynamical Astronomy, **100**, 121-140 (2008).

36. Kramm, U.; Nettelmann, N.; Redmer, R.; Stevenson, D. J.. On the degeneracy of the tidal Love number k2 in multi-layer planetary models: application to Saturn and GJ436b. Astronomy & Astrophysics, **528**, id.A18 (2011).

37. Lammer, H.; Khodachenko, M. L.; Khodachenko, M. L., Herbert, I. M., Lichtenegger, I. M., Kulinov, Y. N.. Impact of Stellar Activity on the Evolution of Planetary Atmospheres and Hability. In: Extrasolar Planets: Formation, detection and Dynamics. Dvorak, R. (Ed.) (2008).

38. Lammer, H.; Bredehft, J. H.; Coustenis, A.; Khodachenko, M. L.; Kaltenegger, L.; Grasset, O.; Prieur, D.; Raulin, F.; Ehrenfreund, P.; Yamauchi, M.; and 7 coauthors. What makes a planet habitable? The Astronomy and Astrophysics Review, **17**, 181-249 (2009).

39. Léger, A.; Rouan, D.; Schneider, J.; Barge, P.; Fridlund, M.; Samuel, B.; Ollivier, M.; Guenther, E.; Deleuil, M.; Deeg, H. J.; and 151 coauthors. Transiting exoplanets from the CoRoT space mission. VIII. CoRoT-7 b: the first super-Earth with measured radius. Astronomy and Astrophysics, **506**, 287-302 (2009).

40. Léger, A.; Grasset, O.; Fegley, B.; Codron, F.; Albarede, A. F.; Barge, P.; Barnes, R.; Cance, P.; Carpy, S.; Catalano, F.; Cavarroc, C.; Demangeon, O.; Ferraz-Mello, S.; Gabor, P.; Griemeier, J.-M.; Leibacher, J.; Libourel, G.; Maurin, A.-S.; Raymond, S. N.; Rouan, D.; Samuel, B.; Schaefer, L.; Schneider, J.; Schuller, P. A.; Selsis, F.; Sotin, C.. The extreme physical properties of the CoRoT-7b super-Earth. Icarus, **213**, 1-11. (2011).

41. Levrard, B., Correia, A. C. M., Chabrier, G., Baraffe, I., Selsis, F., Laskar, J.. Tidal dissipation within hot Jupiters: a new appraisal. Astronomy and Astrophysics, **462**, L5-L8 (2007).

42. Madhusudhan, Nikku; Lee, Kanani K. M.; Mousis, Olivier. A Possible Carbon-rich Interior in Super-Earth 55 Cancri e. The Astrophysical Journal Letters, (**759**), L40 (2012).

43. Mardling, R. A.. Long-term tidal evolution of short-period planets with companions. Monthly Notices of the Royal Astronomical Society, (**382**) 1768-1790 (2007).

44. Matsumura, Soko; Takeda, Genya; Rasio, Frederic A. On the Origins of Eccentric Close-In Planets. The Astrophysical Journal, Volume 686, Issue 1, pp. L29-L32 (2008).

45. Michtchenko, T. A., Ferraz-Mello, S.. Resonant Structure of the outer solar system in the neighborhood of the planets. The Astronomical Journal **122**, 474-481 (2001).

46. Mignard, F.. The evolution of the lunar orbit revisited - I. Moon and Planets, **20**, 301-315 (1979).

47. Muirhead, Philip S.; Johnson, John Asher; Apps, Kevin; Carter, Joshua A.; Morton, Timothy D.; Fabrycky, Daniel C.; Pineda, John Sebastian; Bottom, Michael; Rojas-Ayala, Brbara; Schlawin, Everett; Hamren, Katherine; Covey, Kevin R.; Crepp, Justin R.; Stassun, Keivan G.; Pepper, Joshua; Hebb, Leslie; Kirby, Evan N.; Howard, Andrew W.; Isaacson, Howard T.; Marcy, Geoffrey W.; Levitan, David; Diaz-Santos, Tanio; Armus, Lee; Lloyd, James P.. Characterizing the cool kois. iii. KOI 961: a small star with large proper motion and three small planets. The Astrophysical Journal, Volume 747, Issue 2, article id. 144 (2012).

48. Murray, C. D, Dermott, S. F.. Solar System Dynamics, Cambridge University Press (1999).

49. Rappaport, Nicole; Bertotti, Bruno; Giampieri, Giacomo; Anderson, John D.. Doppler Measurements of the Quadrupole Moments of Titan. Icarus, **126**, 313-323 (1997).

50. Ragozzine, D.; Wolf, A. S.. Probing the Interiors of very Hot Jupiters Using Transit Light Curves. The Astrophysical Journal, **698**, 1778-1794 (2009).

51. Rivera, E. J., Gregory Laughlin, R. Paul Butler, Steven S. Vogt, Nader Haghighipour, Stefano Meschiari. The Lick-Carnegie exoplanet survey: a Uranus-mass fourth planet for GJ 876 in an extrasolar laplace configuration. The Astrophysical Journal, **719**, 890-899 (2010).

52. Rodríguez, A., Ferraz-Mello, S.. Tidal decay and circularization of the orbits of short-period planets. EAS Publications Series, **42**, 411-418 (2010).

53. Rodríguez, A.. Evolução Orbital de Planetas Quentes Atribuída ao Efeito de Maré. PhD Thesis, Universidade de Sao Paulo, Brazil, (2010).

54. Rodríguez, A.; Ferraz-Mello, S.; Michtchenko, T. A.; Beaugé, C.; Miloni, O.. Tidal decay and orbital circularization in close-in two-planet systems. Monthly Notices of the Royal Astronomical Society, 415, 2349-2358 (2011).

55. Rodríguez, A. C., Callegari, N. Jr., Michtchenko, T., Hussmann, H.. Spin-orbit evolution of hot Super-Earths. Monthly Notices of the Royal Astronomical Society, **427**, 2239-2250 (2012).

56. Seager, S.; Hui, Lam. Constraining the Rotation Rate of Transiting Extrasolar Planets by Oblateness Measurements. The Astrophysical Journal, **574**, 1004-1010 (2002).

57. Showman, Adam P.; Polvani, L. M.. Equatorial Superrotation on Hot Jupiters. The Astrophysical Journal, **738**, Issue 1, article id. 71 (2011).

58. Schubert G., Anderson J. D., Spohn T., McKinnon W. B.. Interior composition, structure and dynamics of the Galilean satellites. In: Jupiter. The Planet, Satellites and Magnetosphere, Edited by Fran Bagenal, Timothy E. Dowling, William B. McKinnon, (**1**), 281-306. (2004).

59. Spiegel, David S.; Raymond, Sean N.; Dressing, Courtney D.; Scharf, Caleb A.; Mitchell, Jonathan L.. Generalized Milankovitch Cycles and Long-Term Climatic Habitability. The Astrophysical Journal, **721**, 1308-1318 (2010).

60. Tadeu dos Santos, M.  G. G. Silva  S. Ferraz-Mello  T.A. Michtchenko. A new analysis of the GJ581 extrasolar planetary system. Celestial Mechanics and Dynamical Astronomy, (**113**), 49-62 (2012).

61. Torres, Guillermo; Fressin, Franois; Batalha, Natalie M.; Borucki, William J.; Brown, Timothy M.; Bryson, Stephen T.; Buchhave, Lars A.; Charbonneau, David; Ciardi, David R.; Dunham, Edward W.; and 20 coauthors. Modeling Kepler Transit Light Curves as False Positives: Rejection of Blend Scenarios for Kepler-9, and Validation of Kepler-9 d, A Superearth-size Planet in a Multiple System. The Astrophysical Journal, **727**, Issue 1, article id. 24 (2011).

62. van Hoolst, T.; Rambaux, N.; Karatekin, O.; Dehant, V.; Rivoldini, A.. The librations, shape, and icy shell of Europa. Icarus, Volume 195, Issue 1, p. 386-399 (2008).

63. Valencia, Diana; Sasselov, Dimitar D.; O'Connell, Richard J.. Detailed Models of Super-Earths: How Well Can We Infer Bulk Properties?. The Astrophysical Journal, **665**, 1413-1420 (2007a).

64. Valencia, D., Sasselov, Dimitar D.; O'Connell, Richard J.. Radius and Structure Models of the First Super-Earth Planet. The Astrophysical Journal, **656**, 545-551 (2007b).

65. Valencia, D.. Characterising Super-Earths. EPJ Web of Conferences, **11**, 03001 (2011).

66. Zuluaga, Z. I., Cuartas-Restrepo, P. A.. The role of rotation on the evolution of dynamo generated magnetic fields in Super Earths. Icarus 217, 88-102 (2012).

67. Wisdom, J., Peale, S. J., Mignard, F.. The chaotic rotation of Hyperion. Icarus, **58**, 137-152 (1984).

68. Wisdom, J.. Rotational dynamics of irregularly shaped natural satellites. The Astronomical Journal **94**, 1350-1360 (1987).

69. Wisdom, J.. Spin-Orbit Secondary Resonance Dynamics of Enceladus. The Astronomical Journal **128**, 484-491 (2004).

70. Wisdom, J.. Tidal dissipation at arbitrary eccentricity and obliquity. Icarus, **193**, 637-640 (2008).

## Appendix A: Orbital circularization due to tidal effect

We consider an interacting pair composed of a slow-rotating star and a close-in planet. The aim is to analyze the timescale for orbital circularization due to tidal interaction. We refer to Ferraz-Mello et al. (2008) and Rodríguez and Ferraz-Mello (2010) for assumptions, definitions of quantities and further details.

The average variation of the eccentricity due to the combined effects of stellar and planetary tides is given by

$$\langle \dot{e} \rangle = -\frac{1}{3} n e a^{-5}(18\hat{s} + 7\hat{p}), \tag{9}$$

where

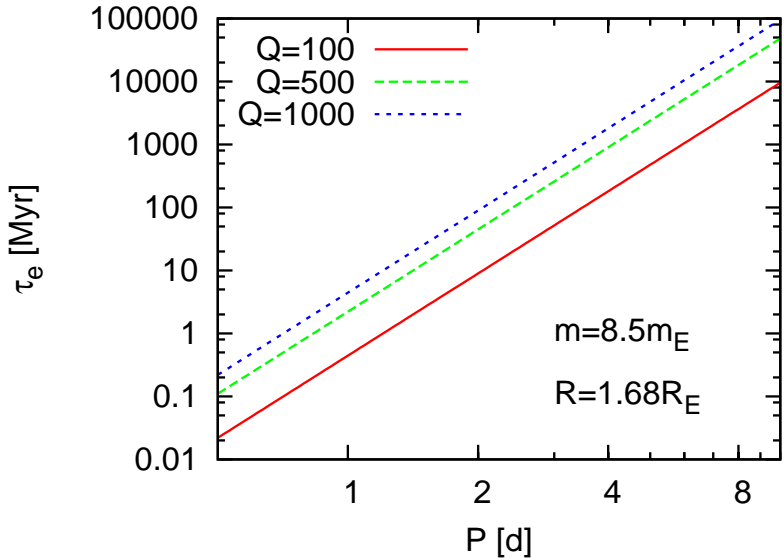

**Fig. 8** Plot of Equation (13) for CoRoT-7b super-Earth planet. The mass of the host star is $m_0 = 0.93 m_{SUN}$.

$$\hat{s} = \frac{9}{4} \frac{k_{20}}{Q_0} \frac{m}{m_0} R_0^5 \quad ; \quad \hat{p} = \frac{9}{2} \frac{k}{Q} \frac{m_0}{m} R^5, \tag{10}$$

are two parameters which stand for stellar and planetary tides, respectively. The symbol 0 refers to the star, $k_2$ is the second degree Love number, $Q$ is the dissipation function or quality factor. It can be shown that the orbital circularization can be accounted by planetary tides only. Indeed, $\hat{s}$ is proportional to $(m/m_0)(1/Q_0)$ which becomes a small quantity for small mass planets and typical stellar $Q_0$ values. Thus, the contribution of stellar tides can be safely neglected in our analysis (see Rodríguez and Ferraz-Mello 2010).

The timescale for orbital circularization can be defined by $\tau_e \equiv e/|\langle \dot{e} \rangle|$, or

$$\tau_e = \frac{3n^{-1}a^5}{7\hat{p}}. \tag{11}$$

Writing Equation (11) as a function of the semi-major axis, we obtain

$$\tau_e = A \, a^{13/2}, \tag{12}$$

where $A = \frac{3\hat{p}^{-1}}{7\sqrt{Gm_0}}$. Alternatively, we can express the result as a function of the orbital period $P$ as follows

$$\tau_e = B \, P^{13/3}, \tag{13}$$

where $B = A \left( \frac{Gm_0}{4\pi^2} \right)^{13/6}$. Figure 8 shows the plot of Equation (13) for a planet with the properties of CoRoT-7b, a Super-Earth planet with $m = 8.5 m_E$ (Ferraz-Mello et al. 2011), $R = 1.68 R_E$, assuming the values $Q = 100$, $Q = 500$, $Q = 1000$, and $k_2 = 0.35$.

We clearly see that the circularization timescale decreases for short-period planets, as Equation (13) indicates. Moreover, noting that $\tau_e \propto Q$, the orbital circularization would become faster in the case of small $Q$ values (i.e., large dissipation). As an example, for $P = 4$ days and $Q = 100$ we have $\tau_e \simeq 181$ Myr.

It is important to note that the parameter $B$ is linearly dependent on $Q/k_2$, which is a quantity poorly known for extrasolar planets. Hence, the plot shown in Figure 8 can be strongly modified if other values of the planet dissipation are considered.

The result (13) can be useful for quantifying the efficiency of tides to produce orbital circularization of close-in planets. Note that, in some cases, $\tau_c$ can be compared with the age of the system, indicating that the orbit of the close-in planet should be circularized during the planet lifetime. However, for large $P$, $\tau_c$ can be even larger than the age of a typical planetary system, in which case a non-circular orbit should be expected due to tidal interaction.

## Appendix B: Capture in spin-orbit resonance

The numerical exploration of the planet rotation, which is subject to the gravitational torque of the star, has shown different behaviors, including the oscillation around spin-orbit resonances. When a dissipative effect like the tidal torque is included, the rotation can be captured in a resonant motion. The specific capture depends on the eccentricity, and also on $Q$ and $\epsilon$. Hence, as $e$ is tidally damped, the capture should become unstable and the rotation can achieve another resonant state, which, at the same time, should result in a temporary trapping. When the orbit completes the circularization due to the tidal torque, the final evolution results in synchronization of the orbital and rotational periods (i.e., the 1:1 spin-orbit resonance). The reader is referred to Goldreich and Peale (1966) and Rodríguez et al. (2012) for further details on the spin-orbit evolution of close-in planets.

Let us first consider the torque due to the prolateness or permanent equatorial deformation (i.e., $a \neq b$) on a rotating body of mass $m$ and radius $R$. The maximum torque, averaged over an orbital period, is given by

$$< N > = -\frac{3}{2} n^2 (B - A) H(p, e) \tag{14}$$

(see Goldreich and Peale, 1966 and Equation (1)), where $H(p, e)$ are power series in $e$ and $p = \Omega/n$, with $\Omega$ the angular velocity of rotation of the deformed body (Goldreich and Peale 1966). Equation (14) assumes that there is commensurability between $\Omega$ and $n$, indicating that $p = \ldots - 1, -1/2, 1, 1/2 \ldots$

In addition to the above torque, we also consider the tidal torque driven by the central body of mass $m_0$. The average tidal torque reads

$$< T > = -\frac{3k_2 G m_0^2 R^5}{8a^6} [4\varepsilon_0 + e^2(-20\varepsilon_0 + 49\varepsilon_1 + \varepsilon_2)] \tag{15}$$

(see Ferraz-Mello et al. 2008), where $k_2$ is the second degree Love number, whereas $\varepsilon_i$ are the phase lags of tidal waves with frequency $\nu_i$. The phase lags account for the internal viscosity, which introduces a delay between the action of the tidal force and the corresponding deformation.

Several tidal models can be used to fix the dependence between phase lags and frequencies, that is, the function $\varepsilon_i(\nu_i)$. We first consider what is usually referred to the linear model, where $\varepsilon_i = \nu_i \Delta t$, where $\Delta t$ is known as time lag and is considered constant in the linear model (Mignard, 1979). The frequencies associated with the phase lags appearing in Equation (15) are $\nu_0 = 2\Omega - 2n$, $\nu_1 = 2\Omega - 3n$ and $\nu_2 = 2\Omega - n$ (see Ferraz-Mello et al. 2008). Replacing these values in Equation (15) and applying the linear model, we obtain

$$< T > = -\frac{3k_2 \Delta t G m_0^2 R^5 n}{2a^6} [2p - 2 + (15p - 27)e^2]. \tag{16}$$

The time lag can be related to the most used quantity $Q$, the dissipation function. Since $\varepsilon_i \simeq Q_i^{-1}$, it follows, under the assumption of a linear model, $Q_0^{-1} \simeq 2n\Delta t(p - 1)$, $Q_1^{-1} \simeq n\Delta t(2p - 3)$, $Q_2^{-1} \simeq n\Delta t(2p - 1)$. We note that singularities in $Q_i$ are associated with spin-orbit commensurability with $p = 1$, $p = 3/2$ and $p = 1/2$. Since we restrict our investigation to the cases of $1 : 1$ and $3 : 2$ spin-orbit resonances, we call $Q = Q_2$ in order to avoid $Q_i$-singularities. Hence, the relationship between $Q$ and $\Delta t$ is given by $Q = 1/[n\Delta t(2p - 1)]$.

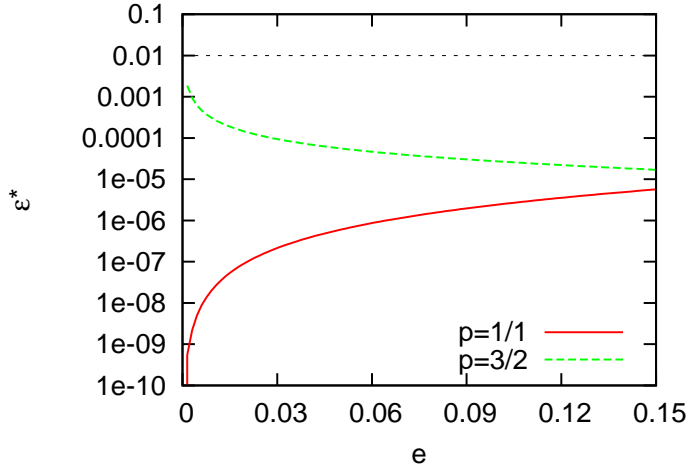

**Fig. 9** Critical value of the equatorial ellipticity as a function of orbital eccentricity for two spin-orbit resonances. The CoRoT-7+7b system illustrates the example.

### Stationary solutions

By definition, the stationary solutions of the rotation are those which satisfy

$$< N > + < T > = 0. \tag{17}$$

Because we are interested in those solutions for which there exists commensurability between spin and orbital revolutions, we can determine a critical value of $\epsilon$ which allows a spin-orbit resonance motion to be maintained when the planet's rotation is under simultaneous action of two torques. Thus, using Equation (14,16,17) we obtain

$$\epsilon^* = -\frac{1}{\xi}\frac{k_2}{Q}\frac{m_0}{m}\left(\frac{R}{a}\right)^3\frac{[2p-2+(15p-27)e^2]}{(2p-1)H(p,e)}. \tag{18}$$

where we have used the third Kepler law and $(B-A) \simeq C\epsilon = \xi m R^2 \epsilon$, where $C$ is the moment of inertia about the rotation axis and $\xi = \frac{C}{mR^2}$, $0 < \xi \leq 2/5$. ($\epsilon^*$ must not be confused with that given in Equation (8)).

The condition $\epsilon > \epsilon^*$ is usually known as stability condition of the $p$-resonance (e.g. Goldreich, 1966). In fact, the stability condition requires that $< T >$ not exceed the maximum restoring torque $< N >$ and, for that reason, $\epsilon^*$ should be considered as a critical value.

We note that the case $p = 1$ is in agreement with the result found in Ferraz-Mello et al. (2008) for the synchronous motion[16].

Figure 9 shows the variation of $\epsilon^*$ with the orbital eccentricity, taking the planet CoRoT-7b as an example. We adopt $k_2 = 0.35$, $\xi = 0.4$ and $Q = 100$. The cases of $1:1$ (synchronous rotation) and $3:2$ spin-orbit resonances are shown. For second order in eccentricity, we have $H(1,e) = 1 - 5e^2/2$ and $H(3/2,e) = 7e/2$. The dashed horizontal line indicates the value of $\epsilon$ used in the simulations for CoRoT-7b ($\epsilon = 0.00992$; Figures 2(a,b)). We note that $\epsilon > \epsilon^*$ in both cases, indicates that the resonant motion should be stable for the range of considered eccentricity. However, $\epsilon^*$ can reach high values for very small $e$, as Equation (18) indicates. On the other hand, we have seen in the numerical simulations (assuming no tides, Figure 2(b)) that the domain of the $3 : 2$ spin-orbit resonance is very small for almost-circular orbits.

---

[16] For comparison it is necessary to consider the relationship between $J_{22}$ (the equatorial ellipticity) and $B - A$, that is, $B - A = 4C_{22}mR^2$ (see Beutler 2005).